\documentclass[a4paper,11pt]{article}

\usepackage{amsmath,amssymb}    
\usepackage{color}
\usepackage{graphicx}
\usepackage{hyperref}            
\usepackage{multirow,makecell}   
\usepackage{float}
\usepackage{tikz}
\usetikzlibrary{calc}\usetikzlibrary{decorations.markings,arrows}
\usepackage{cite}  
\usepackage[utf8]{inputenc}
\numberwithin{equation}{section}  

\usepackage{MnSymbol}

\DeclareFixedFont\trfont{OT1}{phv}{b}{sc}{11}

\begin{document}

\thispagestyle{empty}
%\today
\begin{flushright}\footnotesize
	%\texttt{ITEP-TH-nn/yy}\\
	\texttt{NORDITA 2020-101} \\
	
	%\vspace{0.3cm}
\end{flushright}

\renewcommand{\thefootnote}{\fnsymbol{footnote}}
\setcounter{footnote}{0}

\begin{center}
	{\Large\textbf{\mathversion{bold} 
		Wilson loops in circular quiver SCFTs at strong coupling  
		}
		\par}
	
	\vspace{0.5cm}
	
	\textrm{Hao Ouyang}
	\vspace{4mm}
	
	{\small 
		\textit{Nordita, KTH Royal Institute of Technology and Stockholm University\\
			Roslagstullsbacken 23, SE-106 91 Stockholm, Sweden}\\
				\texttt{hao.ouyang@su.se}
		%\vspace{3mm}
	}

	%\vspace{3mm}
	
	%%%%%%%%

	\par\vspace{1.5cm}
	
	\textbf{Abstract} \vspace{3mm}
	
	\begin{minipage}{\textwidth}
		We study circular BPS Wilson loops in the  $\mathcal{N}=2$ superconformal $n$-node quiver theories at
		large $N$ and strong 't Hooft coupling by using localization. We compute the expectation values of Wilson loops in the limit when the 't Hooft couplings are hierarchically different and when they are nearly equal. Based on these results, we make a conjecture for arbitrary strong couplings. 
	\end{minipage}
	
\end{center}

\vspace{1.5cm}

\newpage
\tableofcontents

\section{Introduction}
Testing holographic duality usually involves non-perturbative aspects of interacting field theory.
The $\mathcal{N}=4$ super-Yang-Mills (SYM) theory offers an ideal playground where many observables can be obtained analytically beyond the perturbation theory.
One such observable of interest is the BPS Wilson loop.
The expectation value of the BPS Wilson loop in  $\mathcal{N}=4$ SYM can be reduced to the computation of a Gaussian matrix model \cite{Erickson:2000af,Drukker:2000rr}, which can be derived by using the supersymmetric localization technique \cite{Pestun:2007rz}.
This exceptional result provides a nontrivial consistency check for the duality relation between the Wilson loop  expectation value and the string worldsheet path integral \cite{Maldacena:1998im,Rey:1998ik}.

\begin{figure}[H]
	\centering
	\begin{tikzpicture}[thick, scale=0.8]
		\tikzset{->-/.style={decoration={
					markings,
					mark=at position .5 with {\arrow{>}}},postaction={decorate}}}
		% All nodes, node labels, and loops
		\foreach \ang\lab\anch in {90/1/north, 45/2/{north east}, 0/3/east, 270/i/south, 180/{n-1}/west, 135/n/{north west}}{
			\draw ($(0,0)+(\ang:3)$) circle (.3);
			\node at ($(0,0)+(\ang:3)$) {$N$};
			\node[anchor=\anch] at ($(0,0)+(\ang:2.8)$) {$\lambda_{\lab}$};
			
		}
		
		% Top part of circle, arrows between different nodes and their labels
		\foreach \ang\lab in {90/1,45/2,180/{n-1},135/n}{
			\draw[->-] ($(0,0)+(\ang-5.5:3)$) arc (\ang-5.5:\ang-39.5:3);
					
		}
		
		% Bottom part of circle, arrows between different nodes and their labels
		\draw[->-] ($(0,0)+(-5.5:3)$) arc (354.5:325+5.5:3);
		\draw[->-] ($(0,0)+(305-5.5:3)$) arc (305-5.5:270+5.5:3);
		\draw[->-] ($(0,0)+(270-5.5:3)$) arc (270-5.5:235+5.5:3);
		\draw[->-] ($(0,0)+(215-5.5:3)$) arc (215-5.5:180+5.5:3);

		% Ellipsis
		\foreach \ang in {310,315,320,220,225,230}{
			\draw[fill=black] ($(0,0)+(\ang:3)$) circle (.02);
		}
		
	\end{tikzpicture}
	\caption{ $n$-node quiver diagram.  Each node represents an $SU(N)$ gauge group factor and the node $n+k$ is identified with the node $k$.
		Each arrow represents a bifundamental hypermultiplet.  }
	\label{fig:quiver}
\end{figure}
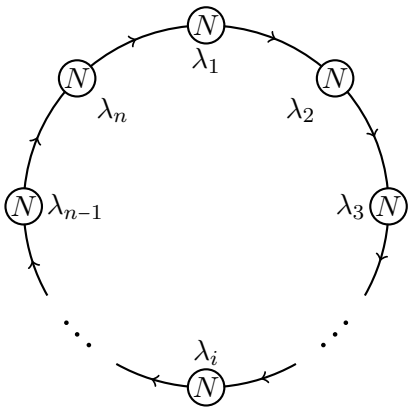

In this work  we study half-BPS circular Wilson loops in $\mathcal{N}=2$ superconformal $A_{n-1}$ quiver gauge theories at strong coupling. There are  $n$ independent 't Hooft couplings for each gauge group factor, which gives rise to interesting dynamics but still allows analytical control.
The field contents are graphically represented in fig. \ref{fig:quiver}. 
When the couplings are equal, they are equivalent to the $\mathbb{Z}_n$ orbifolds of $\mathcal{N}=4$ SYM \cite{Lawrence:1998ja}. Therefore they are probably the next simplest theories after the $\mathcal{N}=4$ SYM.
Localization technique is also applicable in such theories.
Unlike the $\mathcal{N}=4$ SYM, the resulting matrix model is interacting.
Many efforts have been devoted to study these theories by  exploiting the  localization matrix model  \cite{Rey:2010ry,Mitev:2014yba,Mitev:2015oty,Pini:2017ouj,Billo:2018oog,Billo:2019fbi,Zarembo:2020tpf,Fiol:2020ojn,Beccaria:2020hgy}.

These quiver theories admit  a dual holographic description as type IIB string theory on the $AdS_5\times (S^5/\mathbb{Z}_n)$ geometry \cite{Kachru:1998ys}, where $\mathbb{Z}_n$ acts on $\mathbb{R}^4\subset\mathbb{R}^6$ the embedding space of $S^5$.
There are $n-1$ independent collapsed two-cycles hidden in the fixed point.
It is convenient to parameter the 't Hooft couplings as:
\begin{equation}
	\lambda_l=\frac{2\pi\lambda}{n\theta_l},~~~l=1,2,...,n
\end{equation}
where  the $\theta$-parameters are constrained by 
\begin{equation}
	\sum_{l=1}^n\theta_l=2\pi,
\end{equation}
and $\lambda$ is the effective coupling:
\begin{equation}
	\lambda=\Big(\frac{1}{n}\sum_{l=1}^n \lambda_l^{-1}\Big)^{-1}.
\end{equation}
 From the holographic perspective  $\lambda$ is related to  the string tension $T$ via $\lambda=4 \pi^2 T^2$ \cite{Lawrence:1998ja,Klebanov:1999rd,Gadde:2010zi}
and the $\theta$-parameters are proportional to the fluxes of the NSNS $B$-field through  appropriate collapsed two-cycles of the orbifold \cite{Lawrence:1998ja,Klebanov:1999rd}. We are interested in the supergravity limit where $\lambda$ is large with fixed $\theta$-parameters.

At strong coupling the leading order Wilson loop expectation value was obtained in \cite{Rey:2010ry}  which is  essentially the same as that in $\mathcal{N}=4$ SYM and independent of the $\theta$-parameters. The "one-loop" correction beyond the leading exponential in the two-node theory was computed in \cite{Zarembo:2020tpf}. The goal of this paper is to extend the result to the $n$-node theory.
Hopefully, our result will provide important insight into the dynamics of string theory on the orbifold with nontrivial $B$-field fluxes.

This paper is organized as follows. In section \ref{s2} we briefly review Wilson loops at strong coupling in the localization approach. 
We then compute the expectations of the Wilson loops in the cases when the 't Hooft couplings are hierarchically different in section \ref{s3} and nearly equal in section \ref{s4}.
We make a conjecture for arbitrary values of strong 't Hooft couplings in section \ref{s5}.
Conclusions are given in section \ref{sc}.

\section{Setup}\label{s2}
In this section we will investigate Wilson loop at strong coupling in the $n$-node quiver theory following similar analysis as in \cite{Zarembo:2020tpf}. So we will be brief in the description and refer the reader to that reference for all details.

\subsection{Localization}

We consider an $A_{n-1}$ quiver theory  with gauge group $SU(N)^n$.
There are $n$ vector multiplets  and $n$ bifundamental hypermultiplets.
The partition function of the theory of interest on a four-sphere can be
localized to be the matrix model \cite{Pestun:2007rz}:
\begin{equation}\label{localizationMM}
	Z=\int_{}^{}\prod_{l=1}^{n}\prod_{i=1}^{N}da_{li}\,\,
	\frac{\prod_{l}^{}\prod_{i<j}^{}(a_{li}-a_{lj})^2H^2(a_{li}-a_{lj})}{\prod_{l}^{}\prod_{ij}^{}H(a_{li}-a_{l+1\,j})}\,
	\,{e}\,^{-\sum_{l}^{}\frac{8\pi ^2N}{\lambda _l}\sum_{i}^{}a_{li}^2},
\end{equation}
where $H(x)$ is given by
\begin{equation}
	H(x)=\prod_{n=1}^{\infty }\left(1+\frac{x^2}{n^2}\right)^n\,{e}\,^{-\frac{x^2}{n}}.
\end{equation}
We are interested in the expectation value of Wilson loops:
\begin{equation}
	W_l=\left\langle \frac{1}{N}\,{\mathrm P}\exp
	\left[\oint_C ds\,\left(i\dot{x}^\mu A_{l\mu} +|\dot{x}|\Phi _l\right) \right]
	\right\rangle,
\end{equation}
where $C$ is the equatorial circle of the four-sphere. In the matrix model, they can be computed as
\begin{equation}
	W_l=\left\langle \frac{1}{N}\sum_{i}^{}\,{e}\,^{2\pi a_{li}}\right\rangle.
\end{equation}
In the following we will be interested in the large $N$ limit in which the distribution of the eigenvalues $a_{lj}$
is characterized by the densities:
\begin{equation}
	\rho _l(x)=\left\langle \frac{1}{N}\,\sum_{i}^{}\delta (x-a_{li})\right\rangle,
\end{equation}
They satisfy the saddle-point equations \cite{Brezin:1977sv} 
\begin{equation}\label{saddle}
\begin{split}
&\strokedint_{-\mu _l}^{\mu _l}dy\,\rho _l(y)\left(\frac{1}{x-y}-K(x-y)\right)
+\frac{1}{2}\int_{-\mu _{l+1}}^{\mu _{l+1}}dy\,\rho _{l+1}(y)K(x-y)\\&
+\frac{1}{2}\int_{-\mu _{l-1}}^{\mu _{l-1}}dy\,\rho _{l-1}(y)K(x-y)=\frac{8\pi ^2}{\lambda _l}\,x,
\end{split}
\end{equation}
where
\begin{equation}
	K(x)=-\frac{H'(x)}{H(x)}=x\left(\psi (1+ix)+\psi (1-ix)+2\gamma \right).
\end{equation}
The expectation values of Wilson loops are then given by 
\begin{equation}\label{wils-int}
	W_l=\int_{-\mu _l}^{\mu _l}dx\,\rho_l (x)\,{e}\,^{2\pi x}.
\end{equation}
When $\lambda_l=\lambda$ for all $l$, the solution is given by the Wigner distribution
\begin{equation}
\rho_l (x)=	\rho (x)=\frac{2}{\pi \mu ^2}\,\sqrt{\mu ^2-x^2}
\end{equation}
with
\begin{equation}
	\mu =\frac{\sqrt{\lambda }}{2\pi }\,.
\end{equation}
In this case the Wilson loop expectation values are the same as that in the $\mathcal{N}=4$ SYM
\begin{equation}\label{leading}
	W_l=W_{\mathrm{SYM}}=\sqrt{\frac{2}{\pi }} \lambda ^{-\frac{3}{4}} {e} ^{\sqrt{\lambda }}.
\end{equation}
When the 't Hooft couplings are unequal but strong, the bulk of the distributions can still be approximated by the Wigner distribution  \cite{Rey:2010ry,Zarembo:2020tpf}. The leading order Wilson loop expectation values are still given by (\ref{leading}).
The differences of the 't Hooft couplings lead to $O(\lambda^0)$ corrections to the position of the endpoints.
The boundary behavior of the density distributions is complicated and determines the Wilson loop expectation values to the next order in $1/\sqrt{\lambda}$.
To obtain boundary behavior we adopt the method used in the study of the $\mathcal{N}=2^*$ theory \cite{Chen:2014vka}.
We need to solve the distributions in the bulk and then at the boundary and finally match the two solutions.

\subsection{Bulk}
When the effective coupling $\lambda$ is large, $\mu_l$ is of order $\sqrt{\lambda}$.
Therefore one can use the large-distance asymptotics approximation of the kernel $K$ in the integral equations:
\begin{equation}
	K(x)\simeq x\log x^2+2\gamma x+\frac{1}{6x}\equiv K^\infty (x).
\end{equation}
The observation in \cite{Zarembo:2020tpf} 
\begin{equation}
	\begin{split}
		&\int_{-\mu }^{\mu }dy\,\sqrt{\mu ^2-y^2}\,K^\infty (x-y)
		=\frac{\pi }{3}\,x^3+ \left(\pi \mu ^2\log\frac{\mu \,{e}\,^{\gamma +\frac{1}{2}}}{2}+\frac{\pi }{6}\right)x
		\nonumber \\
		&\int_{-\mu }^{\mu }dy\,\,\frac{K^\infty (x-y)}{\sqrt{\mu ^2-y^2}}=2\pi x\log\frac{\mu \,{e}\,^{\gamma +1}}{2}
		\nonumber \\
		&
		\int_{-\mu }^{\mu }dy\,\,
		\frac{K^\infty (x-y)}{\left(\mu ^2-y^2\right)^{n+\frac{1}{2}}}
		=-\frac{2^n(n-1)!\pi }{(2n-1)!!\mu ^{2n}}\,x,\qquad n=1,2,\ldots 		
	\end{split}
\end{equation}
suggests that strong-coupling expansions of the densities take the following form:
\begin{equation}\label{ansatz-bulk}
	\rho _l(x)=A\sqrt{\mu_l ^2-x^2}+\frac{2\mu _lAB_l}{\sqrt{\mu_l ^2-x^2}}
	+\frac{4\mu _l^2AC_l}{\left(\mu _l^2-x^2\right)^{\frac{3}{2}}}+\ldots .
\end{equation}
The unit normalization of the densities gives 
\begin{equation}\label{Bl}
	B_l=\frac{1}{2\pi A\mu _l}-\frac{\mu _l}{4}\,.
\end{equation}
Plugging the asymptotic kernel into the saddle-point equations (\ref{saddle}), we find
\begin{equation}
\begin{split}
	\label{ABC-eq}
	&1-\mu _{l}^2\log\frac{\mu _{l}\,{e}\,^{\gamma +\frac{1}{2}}}{2}-4B_{l}\mu _{l}\log\frac{\mu _{l}\,{e}\,^{\gamma +1}}{2}+8C_{l}	
	\\ &
	+\frac{1}{2}\mu _{l-1}^2\log\frac{\mu _{l-1}\,{e}\,^{\gamma +\frac{1}{2}}}{2}
	+2B_{l-1}\mu _{l-1}\log\frac{\mu _{l-1}\,{e}\,^{\gamma +1}}{2}
	-4C_{l-1}
	 \\ &
	+\frac{1}{2}\mu _{l+1}^2\log\frac{\mu _{l+1}\,{e}\,^{\gamma +\frac{1}{2}}}{2}
	+2B_{l+1}\mu _{l+1}\log\frac{\mu _{l+1}\,{e}\,^{\gamma +1}}{2}
	-4C_{l+1}=\frac{8\pi }{A\lambda _{l}}.	
\end{split}
\end{equation}
The sum of the $n$ equations determines $A$:
\begin{equation}\label{alam}
	A=\sum_{l=1}^n\frac{8\pi }{n\lambda _l}=\frac{8\pi }{\lambda} .
\end{equation}
Then the normalization (\ref{Bl}) leads to
\begin{equation}\label{mub}
\mu _l=\frac{\sqrt{16 \pi ^2 B_l^2+\lambda }-4 \pi  B_l}{2 \pi }=\frac{\sqrt{\lambda }}{2 \pi }-2 B_l+O\left(\sqrt{\frac{1}{\lambda }}\right),
\end{equation}
because $B_l$ should stay finite in the limit of large $\lambda $ .
Therefore the shifts of the endpoints from the Gaussian-model prediction $\mu =\sqrt{\lambda }/2\pi $ at strong coupling are
\begin{equation}
	\Delta_j=\mu-\mu_j=-2B_j.
\end{equation}

Substituting (\ref{alam}) and (\ref{mub}) into (\ref{ABC-eq}), the difference between the $l$th and $(l+1)$th equation in the large $\lambda $ limit  imposes the constraint
\begin{equation}\label{bc}
	2 B_{l-1}^2-6 B_{l}^2+6 B_{l+1}^2-2 B_{l+2}^2+4 C_{l-1}-12 C_{l}+12 C_{l+1}-4 C_{l+2}+\frac{n\theta_{l}}{2\pi}-\frac{n\theta_{l+1}}{2\pi}=0.
\end{equation}
Therefore we have $n-1$ constraints on the $2 n$ coefficients $B_l$ and $C_l$.
The remaining unknowns are fixed by the boundary behavior.

\subsection{Boundary}
To describe the boundary behavior of the densities, we define the scaling functions
\begin{equation}\label{f}
	f _l(\xi)=\lim_{\lambda\rightarrow\infty} \frac{1}{A\sqrt{2\mu _l}}\rho_l(\mu _l-\xi).
\end{equation}
At large distance $\xi \rightarrow \infty$ they should match the bulk solution (\ref{ansatz-bulk}):
\begin{equation}
	f_l (\xi )=f_l^\infty (\xi )+O(\xi ^{-\frac{5}{2}}),~~~
	 f_l^\infty (\xi )\equiv \sqrt{\xi }+\frac{B_l}{\sqrt{\xi }}+\frac{C_l}{\xi ^{\frac{3}{2}}}.
\end{equation}
To derive the integral equations for $f$ from the saddle-point equations (\ref{saddle}), one can use the trick in \cite{Chen:2014vka} to extract the contribution from the boundary region. 
Equations (\ref{saddle}) can be written as
\begin{equation}\label{rr}
	R_{jl}*\rho _l=\frac{8\pi ^2}{\lambda _j}\,x,
\end{equation}
where $*$ denotes convolution.
 The perturbative bulk solution satisfies an exact equation:
\begin{equation}\label{rri}
	R^\infty _{jl}*\rho^\infty  _l=\frac{8\pi ^2}{\lambda _j}\,x,
\end{equation}
where $R^\infty $ is $R$  with $K$ replaced by $K^\infty $ and $\rho^\infty $ is given by the first three terms of the right hand side of (\ref{ansatz-bulk}). 
Subtracting (\ref{rri}) from (\ref{rr}), we get
\begin{equation}
	R_{jl}*(\rho _l-\rho ^\infty _l)=(R^\infty_{jl} -R_{jl})*\rho ^\infty _l.
\end{equation}
To obtain the integral equations for the scaling functions, we take $x=\mu_j -\xi $ and $y=\mu_l -\eta $. For $\xi\sim O(1)$, we find only $\eta\sim O(1)$ contributes because the integrand on the left hand side decay as $\eta^{-3/2}\log \eta$ away from the boundary and that on the right hand side as $\eta^{-3/2}$.
Therefore convolution integrals can be extended to infinity and we get an integral equation  of the Wiener-Hopf type:
\begin{equation}\label{boundaryinteq}
	\int_{0}^{\infty }R_{jl}(\xi -\eta )\left(f_l(\eta )-f^\infty _l(\eta )\right)
	=\int_{0}^{\infty }\left(R^\infty _{jl}(\xi -\eta )-R_{jl}(\xi -\eta )\right)f_l^\infty (\eta ),
\end{equation}
where the explicit form of the kernel is
\begin{equation}
	R_{jl}(\xi )=\left(\frac{1}{\xi }-K(\xi)\right) \delta_{jl}+K(\xi -\Delta_j+\Delta_l )\frac{\delta_{j\,l-1}+\delta_{j\,l+1}}{2},
\end{equation}
and the same for $R^\infty $ with $K$ replaced by $ K^\infty $. 
The shift of the argument in $K$ is a result of different positions of the end points.
It is advantageous   to work in the Fourier space. The Fourier image
\begin{equation}
	\hat f_l(\omega )=\int_{-\infty }^{+\infty }d\xi \,{e}\,^{i\omega \xi }f_l(\xi ).
\end{equation}
is analytic in the upper half plane of $\omega $. 
The equation (\ref{boundaryinteq}) only holds for $\xi>0$, so we have to introduce an unknown function $X$ which is nonzero for $\xi<0$.
After taking the Fourier transform, convolutions in configuration space become products:
\begin{equation}\label{WHF}
	\hat R( \hat f-\hat f^\infty )=(\hat R^\infty -\hat R)\hat f^\infty +\hat X,
\end{equation}
where $\hat X$ is a negative-half-plane analytic function of $\omega $.
To simplify expression, we denote
\begin{equation}
	\omega=-i\pi u.
\end{equation}
Then $\hat f$ and $\hat X$ are analytic in the left and right half-plane of $u $ respectively.
The explicit expressions for Fourier images appearing in (\ref{WHF}) are
\begin{align}
	\hat R&=i\frac{2 \pi  \sqrt{ -\pi  u+\epsilon } \Gamma \left(-\frac{u}{2}\right) \Gamma \left(1+\frac{u}{2}\right)}{\sqrt{ \pi  u+\epsilon } \Gamma \left(\frac{1}{2}-\frac{u}{2}\right) \Gamma \left(\frac{1}{2}+\frac{u}{2}\right)}\tilde R,\\
	\tilde R_{ij}(u )&=i\delta _{ij}\cot \pi u -\frac{i}{2} \csc \pi u\left(\delta _{i\,j+1}+\delta _{i+1\,j}\right)
	e^{u (\pi  \Delta_i-\pi  \Delta_j)},\\	
	\hat R^\infty &=-\frac{\sqrt{ -\pi  u+\epsilon}}{3 \pi  u^2 \sqrt{ \pi  u+\epsilon }}
	\left(\delta _{ij}(-12 + 5 \pi^2 u^2) +\frac{(12 + \pi^2 u^2)}{2}\left(\delta _{i\,j+1}+\delta _{i+1\,j}\right)
	\left(\frac{\eta_j}{\eta_i}\right)^u \right),\\
	\hat f^\infty_i=&
	\frac{ \sqrt{\pi } \left(1-2 \pi  B_i u-4 \pi ^2 C_i u^2\right)}{2 (\epsilon-\pi  u)^{3/2}},
\end{align}
where a small $\epsilon>0$ is used to express the analytic form of $\mathop{\mathrm{sign}}\omega $:
\begin{equation}
	\mathop{\mathrm{sign}}\omega =\lim_{\epsilon \rightarrow 0}
	\frac{\sqrt{\omega +i\epsilon }}{\sqrt{\omega -i\epsilon }}
	=i\lim_{\epsilon \rightarrow 0}\frac{ \sqrt{ -\pi  u+\epsilon }}{\sqrt{ \pi  u+\epsilon }}.
\end{equation}
The densities and scaling functions should vanish like a square root at the boundary, so we require
\begin{equation}
	\hat f_l(u )=\int_{0 }^{+\infty }d\xi\,{e}\,^{\pi u \xi }f_l(\xi )\sim O((-  u)^{-3/2}),~~~u\rightarrow -\infty.	
\end{equation}
The Taylor expansion of the scaling functions at $u=0$ are determined by
\begin{equation}
	\frac{2 (\epsilon-\pi  u)^{3/2}}{\sqrt{\pi }} \hat f_l\sim
1-2 \pi  B_l u-4 \pi ^2 C_l u^2+O(u^3),~~~u\rightarrow 0,
\end{equation}
because they should match the bulk solution at large $\xi$.

In \cite{Zarembo:2020tpf} the Wiener-Hopf problem (\ref{WHF}) for $n=2$ was reduced to the analytic factorization of the kernel $R$, and  $f_l$ can be expressed through the Wiener-Hopf factors of $R$ via contour integration.
We will use a more direct approach to express $f_l$ here.
Let us first consider an auxiliary problem:
\begin{equation}\label{sfx}
	\tilde R \tilde f=\tilde X,
\end{equation}
where $\tilde X(u)$ is analytic in the  right half-plane and  $\tilde f(u)$ is analytic in the  left half-plane except at $u=0$.
The asymptotic behavior of $\tilde f_l$ at small and large $u$ is required to be
\begin{align}
	\tilde f_l&\stackrel{u\rightarrow 0}{\simeq } -\frac{1}{u}+(2\pi B_l +\log 2)+\frac{1}{24} u \left(-24 \pi  B_l \log 4+\pi ^2 (96 C_l+1)-3 \log ^2 4\right)+O(u^2),  \label{zero}\\
	\tilde f_l&\stackrel{u\rightarrow \infty}{\simeq } O((-u)^{-1/2}). \label{infinity}
\end{align}
Then $\hat f$ can be obtained as
\begin{equation}
	\hat f=\frac{ \Gamma \left(\frac{1}{2}-\frac{u}{2}\right)}{(\epsilon -\pi  u)^{3/2} \Gamma \left(-\frac{u}{2}\right)}
	\tilde f.
\end{equation}
One can check that this is a solution to the Wiener-Hopf problem (\ref{WHF}) with appropriate $\hat X$ and satisfies the required asymptotic conditions at small and large $u$.

We now turn to Wilson loops. One can replace the density in equation (\ref{wils-int}) by the scaling function and extend the integration to $\xi\rightarrow\infty$ because the contribution from the large $\xi$ region is suppressed exponentially. Therefore the strong-coupling expectation values of Wilson loops can be computed as
\begin{equation}
	\begin{split}
	W_l\simeq &A\sqrt{2\mu _l}\,\,{e}\,^{2\pi \mu _l}\int_{0}^{\infty }
d\xi \,f_l(\xi )\,{e}\,^{-2\pi \xi }\\
\simeq&\sqrt{\frac{2}{\pi }}\lambda ^{-3/4}  e^{2\pi \mu_l}\tilde f_l(-2).		
	\end{split}
\end{equation}
We are interested in the Wilson loop expectation value $W_l$ normalized by that in the $\mathcal{N}=4$ SYM at strong coupling:
\begin{equation}\label{ratiow}
	w_l=\lim_{\lambda\rightarrow\infty }\frac{W_l}{W_{\mathrm{SYM}}}=e^{-4\pi B_l}\tilde f_l(-2).
\end{equation}
This quantity is useful for matching gauge theory and string theory results.
In the string calculation there are subtleties related to the string path integral measure.
To avoid this problem, it is convenient to consider the ratio of  Wilson loops
\cite{Forini:2015bgo,Faraggi:2016ekd}.

 When $n=2$, it was observed in \cite{Zarembo:2020tpf} that the problem (\ref{sfx}) is related to  the scattering
theory of the P\"{o}schl-Teller potential and therefore can be solved exactly.
However, it is difficult to find an analogous relation for general $n$.
In the following two sections, we will solve the problem in the limit when the couplings are hierarchically different or nearly equal. Based on these results, we can make a conjecture for the ratios $w_l$ with arbitrary $\theta$-parameters.

\section{Hierarchical couplings}\label{s3}
In this section we compute $f_l$ in the limit $\Delta_1\gg\Delta_2\gg ...\gg\Delta_n$.
We find that in this limit the 't Hooft couplings have a hierarchy, i.e., $\lambda_1\gg \lambda_2\gg...\gg\lambda_n$.

\subsection{Leading order}
Motivated by the $n=2$ result in \cite{Zarembo:2020tpf}, we propose an ansatz when $|\Delta_l-\Delta_j|\gg 1$ for any $l\neq j$
\begin{equation}
	\tilde X_j=a^-_j+\sum_{\substack{l\\\eta_l<\eta_j}} b^{-}_{lj}\left(\frac{\eta_l}{\eta_j}\right)^u ,~~~
	\tilde f_j=a^+_j+\sum_{\substack{l\\\eta_l>\eta_j}} b^{+}_{lj}\left(\frac{\eta_l}{\eta_j}\right)^u\label{tf},
\end{equation}
where  $a^{\pm}_j$ and $b^{\pm}_{lj}$ depend on $u$ and can be written as fractional power series in the ratios of the $\eta$-parameters  defined as
\begin{equation}
	\eta_l=e^{-\pi \Delta_l}.
\end{equation}
Plugging this ansatz into the problem (\ref{sfx}), we find
\begin{equation}
	a_l^{+} S_{lm}+\sum_{\substack{j\\\eta_l>\eta_j}}b^{+}_{lj}S_{jm}=0,
\end{equation}
for any $l$ and $m$ satisfying $\eta_l>\eta_m$. Solving $b_{lj}^{\pm}$ and $a_{l}^{-}$ in terms of $a_{l}^{+}$.
\begin{align}
	b_{lj}^{+}=&-a_l^{+}\sum_{\substack{m\\\eta_m<\eta_l}}S_{lm}S^{(l)-1}_{mj},\\
	a_{l}^{-}=&-a_l^{+}\Big( S_{ll}-\sum_{\substack{m,j\\\eta_l>\eta_j,\eta_l>\eta_m}}S_{lm}S^{(l)-1}_{mj}S_{jl}\Big),\label{al}\\
	b_{lk}^{-}=&-a_l^{+}\Big( S_{lk}-\sum_{\substack{m,j\\\eta_l>\eta_j,\eta_l>\eta_m}}S_{lm}S^{(l)-1}_{mj}S_{jk}\Big),
\end{align}
where $S^{(l)}$ is a submatrix of $S$ containing elements with indices belonging to $\{m\in \mathbb{Z}_n|\eta_l>\eta_m\}$. 
For concreteness we concentrate on the case when $\eta_1\ll\eta_2\ll ...\ll\eta_n$. We denote $\gamma_i\equiv\eta_i/ \eta_{i+1}\ll 1$ for $i=1,...n-1$. 
Writing explicitly these equations for the first few integers $n$, we find

\begin{align}
	a_{l}^{-}=&\frac{i \sin (\pi  (l+1) u)}{2 (\sin (\pi  u) \sin (\pi  l u))}a_l^{+},~~~
	l\neq n,\label{eqal}\\
	a_{n}^{-}=&-i \tan \left(\frac{\pi  n u}{2}\right)a_n^{+},\label{eqan} \\
	b_{lj}^{+}=&{\sin (\pi  j u)}{\csc (\pi  l u)}a_l^{+},~~~1\leqslant j<l<n\\
	b_{nj}^{+}=&{\cos \left(\frac{1}{2} \pi  n u-\pi  j u\right)}{\sec \left(\frac{1}{2} \pi  n u\right)}a_n^{+},~~~1\leqslant j<n\\
	b_{ll+1}^{-}=&-{\sin (\pi  l u)}{\csc (\pi  (l+1) u)} a_l^{-},~~~
	1\leqslant l<n-1\\
	b_{ln}^{-}=&-{\sin (\pi   u)}{}\csc (\pi  (l+1) u) a_l^{-},~~~1\leqslant l<n-1\\
	b_{n-1\,n}^{-}=&-{\cos \left(\pi  \left(\frac{n}{2}-1\right) u\right)}\sec \left(\frac{\pi  n u}{2}\right)a_{n-1}^{-}.
\end{align}
All other $b^\pm_{jk}$ vanish. Equations (\ref{eqal}) and (\ref{eqan}) can be solved by
\begin{equation}
	\begin{split}
	a_{l}^{+}(u)=&\frac{4  \sqrt{\pi } l^{-l u} (l+1)^{(l+1) u} \Gamma (-l u-u-1)}{\Gamma (-u) \Gamma (-l u)} c_l(u),~~~
l\neq n\\
a_{l}^{-}(u)=&\frac{2i l^{-l u} (l+1)^{(l+1) u} \Gamma (1+u) \Gamma (1+l u)}{\sqrt{\pi } \Gamma (2+(l+1) u)} c_l(u),~~~
l\neq n\\
a_{n}^{+}(u)=&\frac{ \Gamma \left(\frac{-n u}{2}\right)}{\Gamma \left(\frac{1}{2} (-n u+1)\right)} c_n(u),~~~
a_{n}^{-}(u)=\frac{i\Gamma \left(\frac{1}{2}+\frac{n u}{2}\right)}{\Gamma \left(1+\frac{n u}{2}\right)} c_n(u),\label{eqan2}		
	\end{split}
\end{equation}
where we have chosen the coefficients in front of $c$ such that
\begin{equation}
	\frac{a^{\pm}_l}{c _l}\stackrel{u\rightarrow \mp\infty}{\sim} 
	O((\mp u)^{1/2}).
\end{equation}
When the $\gamma$-parameters are small, singularities of $\tilde f_j$ ($\tilde X_j$) at large real negative (positive) values of $u$ are suppressed by the
$({\eta_l}/{\eta_j})^u$ factors.
The coefficients $c_j(u)$ are independent of $u$ in the limit $\gamma_j\rightarrow 0$ , $j=1,2,...,n-1$. In principle, they can be solved perturbatively in small $\gamma_{l}$ using the analytic properties of $\tilde f$ and  $\tilde X$. We will show how to do that in the $n=3$ case, but at this moment let us focus on the leading order contribution. Denoting $c_{j}^{(0)}$  the leading order of $c_j(u)$, The condition that singularity of $\tilde f_j$ at $u=-(j+1)^{-1}$, $j=1,2,...,n-1$ is removable leads to
\begin{equation}
	\begin{split}
			&(l+1)^2 (l+2)^{-\frac{l+2}{l+1}} c_{l+1}^{(0)} \gamma_l^{\frac{1}{l+1}}-l^{\frac{l}{l+1}} c_{l}^{(0)}=0,~~~1\leqslant l\leqslant n-2\\
		&(n-1)^{1/n} n c_{n}^{(0)} \gamma_{n-1}^{1/n}-2 (n-1) c_{n-1}^{(0)}=0.
	\end{split}
\end{equation}
We get
\begin{equation}\label{largec}
	c_l^{(0)}=\frac{c_n^{(0)}}{2} \left(l^{-\frac{l}{l+1}} (l+1)^{\frac{1}{l}+1} n^{-\frac{1}{n+1}}  \right) \prod _{k=l}^{n-1} (k+1)^{-\frac{2}{k^2+2 k}} \gamma_k^{\frac{1}{k+1}},~~~1\leqslant l\leqslant n-1.
\end{equation}
Expanding $\tilde f$ at small $u$ using (\ref{eqal})-(\ref{eqan2}), we find
\begin{align}
	c_n(0)=&\frac{\sqrt{\pi } n}{2},\label{cn0}\\
	B_l=&\frac{\log \eta_l-\log \eta_n+n \log 2-\log 2}{2 \pi }-\frac{c_n'(0)}{\pi ^{3/2} n},\\	
	C_l=&\frac{1}{96 \pi ^{5/2} n}
	(12 \left(\log  \eta _l-\log \eta _n+n \log  4-\log  4\right) \left(\log \eta _n-\log \eta _l\right)
	\nonumber\\&
	+\sqrt{\pi } n \left(\pi ^2 \left(12 l^2-12 l n+n^2-1\right)-12 (n-1)^2 \log ^2 2\right)
		\nonumber\\&
		+48 c_n'(0) \left(\log \eta _l-\log \eta _n+n \log  2-\log  2\right)-24 c_n''(0))
		+\frac{l }{\pi ^{3/2}}\sum _{j=l}^{n-1} \frac{c_j(0)}{j+1}. \label{Cl}
\end{align}
When $l=n$, the last term in (\ref{Cl}) is an empty sum and hence equal to zero by convention. We will follow the  same convention   below.
The endpoint positions are then given by
\begin{equation}
	\mu_l=\frac{\sqrt{\lambda}}{2\pi}-2B_l=\frac{\sqrt{\lambda}}{2\pi}+\frac{2c_n'(0)}{\pi ^{3/2} n}-\frac{\log \eta_l-\log \eta_n+n \log 2-\log 2}{ \pi },
\end{equation}
which are consistent with $\mu_l-\mu_j=\Delta_l-\Delta_j$.
Solving  the constraints (\ref{bc}) we find
\begin{equation}
	\begin{split}\label{coupling}
		&\theta_l=\frac{8}{n\sqrt{\pi}}c_l(0)-8\frac{l-1}{n l\sqrt{\pi}}c_{l-1}(0),~~~1\leqslant l\leqslant n-1,\\
		&\theta_n=2 \pi-\frac{8 (n-1)}{ n^2\sqrt{\pi} } c_{n-1}(0)+\sum _{l=1}^{n-1} \frac{8 }{(l+1)n\sqrt{\pi} }c_l(0),
	\end{split}
\end{equation}
To compute the ratios between the expectation values of Wilson loops and that in the $\mathcal{N}=4$ SYM using (\ref{ratiow}), we need to evaluate $\tilde f$ at $u=-2$:
\begin{equation}
f_l(-2)=\frac{c_n(-2) \eta _l^2 \Gamma (n)}{\eta _n^2 \Gamma \left(n+\frac{1}{2}\right)}+\sum _{k=l}^{n-1} \frac{8 \sqrt{\pi } k^{2 k} (k+1)^{-2 (k+1)} l c_k(-2) \eta _l^2}{\eta _k^2}.
\end{equation}
Then the ratios are given by
\begin{equation}\label{expwl1}
w_l=4^{1-n} \left(\frac{  \Gamma (n)c_n(-2)}{ \Gamma \left(n+\frac{1}{2}\right)}+\sum _{k=l}^{n-1} \frac{8 \sqrt{\pi } k^{2 k} (k+1)^{-2 (k+1)} l \eta _n^2 c_k(-2)}{\eta _k^2}\right)e^{\frac{4 c'_n(0)}{\sqrt{\pi } n}}.
\end{equation}

When $\gamma_i\ll 1$ for all $l$, using (\ref{coupling}) we find  $\theta_1\ll \theta_2\ll...\ll\theta_n$ or equivalently $\lambda_1\gg \lambda_2\gg...\gg\lambda_n\simeq n \lambda$ and
\begin{equation}
	\begin{split}
	\gamma_l\simeq&l^l (l+1)^{-2 (l+1)} (l+2)^{l+2} \lambda _l^{-l-1} \lambda _{l+1}^{l+1},~~~1\leqslant l\leqslant n-2\\
	\gamma_{n-1}\simeq&(n-1)^{n-1} n^{-2 n} \pi ^n \lambda ^n \lambda _{n-1}^{-n}.	
	\end{split}
\end{equation}
Plugging into (\ref{largec}), the leading terms of $w_l$ are
\begin{equation}
	\begin{split}\label{largew}
		w_l^{(0)}=&2^{3-2 n}l  n \pi ^{2-2 n} \lambda _l^{2 l+1} \lambda _n^{1-2 n}  \prod _{j=l+1}^{n-1} \lambda _j^2 ,~~~1\leqslant l\leqslant n-1,\\
		w_n^{(0)}=&\frac{\Gamma (n) \Gamma (n+1)}{\Gamma (2 n)}.
	\end{split}
\end{equation}
For the first few cases $n\leqslant 5$, we have
\begin{align}
	n=2:&~~~w_l^{(0)}=(\frac{\lambda _1^3}{\pi ^2 \lambda _2^3},\frac{1}{3}),\\
	n=3:&~~~w_l^{(0)}=(\frac{3 \lambda _1^3 \lambda _2^2}{8 \pi ^4 \lambda _3^5},\frac{3 \lambda _2^5}{4 \pi ^4 \lambda _3^5},\frac{1}{10}),\\
	n=4:&~~~w_l^{(0)}=(\frac{\lambda _1^3 \lambda _2^2 \lambda _3^2}{8 \pi ^6 \lambda _4^7},\frac{\lambda _2^5 \lambda _3^2}{4 \pi ^6 \lambda _4^7},\frac{3 \lambda _3^7}{8 \pi ^6 \lambda _4^7},\frac{1}{35}),\\
	n=5:&~~~w_l^{(0)}=(\frac{5 \lambda _1^3 \lambda _2^2 \lambda _3^2 \lambda _4^2}{128 \pi ^8 \lambda _5^9},\frac{5 \lambda _2^5 \lambda _3^2 \lambda _4^2}{64 \pi ^8 \lambda _5^9},\frac{15 \lambda _3^7 \lambda _4^2}{128 \pi ^8 \lambda _5^9},\frac{5 \lambda _4^9}{32 \pi ^8 \lambda _5^9},\frac{1}{126}).
\end{align}
The first node has the largest 't Hooft couplings and $w_1$ diverges as $\lambda_1^3$.
This limit can be considered as the supergravity decouple limit discussed in \cite{Zarembo:2020tpf}. 
This is different from the true decoupling limit where $\lambda_l\rightarrow 0$ for $l\neq 1$ and the theory becomes equivalent to the super-QCD.
In the true decoupling limit, there should be cubic  as well as  log behavior \cite{Passerini:2011fe}.

In the rest of this section, we will consider some simple cases where it is easy to compute higher order corrections.

\subsection{Higher order corrections in $\gamma_{n-1}$}\label{ss32}
It is a relatively easy task to consider higher order corrections in $\gamma_{n-1}$. When $\gamma_{n-1}$ is finite but the rest of the $\gamma$-parameters are small, the couplings satisfy $\lambda_1\gg \lambda_2\gg...\gg\lambda_{n-2}\gg\lambda$ and $\lambda_{n-1,n}\sim O(\lambda)$.
In this limit $W_j\rightarrow\infty$ and $c_j\rightarrow 0$ for $j=1,2,...,n-2$. We are going to compute the Wilson loop expectation values associated with the last two nodes.
Keeping only the $\gamma_{n-1}$ corrections $\tilde f$ and $\tilde X$ become
\begin{align}
	\tilde{f}&=(a_1^+,...,a_{n-2}^+,
	a_{n-1}^+ +(\sin \pi  u \tan \frac{\pi  n u}{2}+\cos \pi  u)a_{n}^+\gamma_{n-1}^{-u},
	a_{n}^+),\\
	\tilde{X}&=(a_1^-,...,a_{n-2}^-,a_{n-1}^-,a_{n}^--(\sin \pi  u \tan \frac{\pi  n u}{2}+\cos \pi  u) a_{n-1}^-\gamma_{n-1}^{u}).
\end{align}
Focusing on the last two nodes, the factorization problem reduces to
\begin{equation}
	\left(
	\begin{array}{c}
		\tilde X_{n-1}  \\
		\tilde X_{n} \\
	\end{array}
	\right)	=M_n
	\left(
	\begin{array}{c}
		\tilde f_{n-1}  \\
		\tilde f_{n} \\
	\end{array}
	\right),
\end{equation}
with
\begin{equation}\label{Mn}
	M_n=\left(
	\begin{array}{cc}
		i\frac{\sin (\pi  n u)}{2\sin (\pi  u)  \sin (\pi  (n-1) u) }
		& -i \frac{\sin \left(\frac{\pi  n u}{2}\right) \cos \left(\frac{1}{2} \pi  (n-2) u\right) }{\sin (\pi  u) \sin (\pi  (n-1) u)}
		\gamma ^{-n u}\\
		-i \frac{\sin \left(\frac{\pi  n u}{2}\right) \cos \left(\frac{1}{2} \pi  (n-2) u\right) }{\sin (\pi  u) \sin (\pi  (n-1) u)}
		\gamma ^{n u}
		& i\frac{\sin (\pi  n u)}{2\sin (\pi  u)  \sin (\pi  (n-1) u) } \\
	\end{array}
	\right)	,~~~	\gamma=\gamma_{n-1}^{ u/n}.
\end{equation}
The undesired singularities of
\begin{equation}
	\tilde f_{n-1}=\frac{4 \sqrt{\pi } c_{n-1} (n-1)^{u-n u} n^{n u} \Gamma (-n u-1)}{\Gamma (-u) \Gamma (u-n u)}+\frac{c_{n} \gamma ^{-n u} \cos \left(\pi  u-\frac{\pi  n u}{2}\right)\Gamma (-\frac{n u}{2} )}{ \cos \left(\frac{\pi  n u}{2}\right) \Gamma \left(\frac{1}{2}-\frac{n u}{2}\right)},	
\end{equation}
at $u=-(1+2m)/n$, $m=1,2,...$ and
\begin{equation}
	\tilde X_{n}=\frac{i c_{n} \Gamma (\frac{n u+1}{2} )}{\Gamma \left(\frac{n u}{2}+1\right)}-\frac{2 i c_{n-1} (n-1)^{u-n u} n^{n u} \Gamma (u+1) \gamma ^{n u} \cos \left(\pi  u-\frac{\pi  n u}{2}\right)  \Gamma ((n-1) u+1)}{\sqrt{\pi } \cos \left(\frac{\pi  n u}{2}\right)\Gamma (n u+2)},
\end{equation}
at $u=(1+2m)/n$, $m=1,2,...$ 
should be removed. 
Therefore $c_{n-1}$ should have simple poles at $u=-(1+2m)/n$ and $c_{n}$ at $u=(1+2m)/n$.
We make the ansatz
\begin{align}\label{cncn}
	c_n(u)&=K^{-1}+K^{-1}\sum _{m=0}^{\infty } \frac{\sum _{k=0}^{\infty } \gamma ^{2 m+2k+2} g_n(m,k)}{u-(1+2m)/n},\\
	c_{n-1} (u)&=K^{-1}\frac{1}{2} \gamma  (n-1)^{\frac{1}{n}-1} n \left(\frac{1}{n}+u\right)
	\sum _{m=0}^{\infty } \frac{\sum _{k=0}^{\infty } \gamma ^{2 m+2k} g_{n-1}(m,k)}{u+(1+2m)/n}.
\end{align}
where  $g_{n-1,n}(m,k)$ are constants and the overall factor $K$ is determined by (\ref{cn0}).
When the undesired singularities are removed, we find the recurrence relations
\begin{equation}
	\begin{split}\label{rec}
		g_{n-1} (m,n)=&
		\frac{\left(-\frac{1}{4}\right)^m (n-1)^{2 m \left(\frac{1}{n}-1\right)} n^{2 m} \sin \left(\frac{\pi -\pi  m (n-2)}{n}\right) }{\pi  \Gamma (m+1)^2}\\
		&\times  \Gamma (\frac{2 m+1}{n}) \Gamma (\frac{(2 m+1) (n-1)}{n})\left(\delta _{k,0}-\sum _{p=0}^{k-1} \frac{n g_n(p,k-p-1)}{2 (m+p+1)}\right)
		,\\
		g_{n}(m,n)=&\frac{\left(-\frac{1}{4}\right)^m (n-1)^{-\frac{2 (m+1) (n-1)}{n}} n^{2 m} \sin \left(\frac{\pi  (m (n-2)-1)}{n}\right)}{\pi  \Gamma (m+1)^2}\\
		&\times  \Gamma (\frac{2 m+n+1}{n}) \Gamma (\frac{2 m (n-1)+2 n-1}{n})\sum _{p=0}^k \frac{n g_{n-1}(p,k-p)}{2 (m+p+1)}.
	\end{split}
\end{equation}
Using (\ref{coupling}), (\ref{expwl1}) and the recurrence relations, we get
\begin{align}
	\theta_{n-1}=&2 \gamma  (n-1)^{\frac{1}{n}-1} n+\frac{1}{3} \gamma ^3 (n-1)^{\frac{3}{n}-3} n \left(n^2-6 n+6\right)\nonumber\\&
	+\frac{1}{20} \gamma ^5 (n-1)^{\frac{5}{n}-5} n \left(3 n^4-40 n^3+160 n^2-240 n+120\right)+O\left(\gamma ^7\right)\label{thn1}	\\
	w_{n}=&\frac{\sqrt{\pi } 2^{1-2 n} n \Gamma (n)}{\Gamma \left(n+\frac{1}{2}\right)}
	\Big(1+\frac{2 \gamma ^2 (n-1)^{\frac{2}{n}-1} n^2}{2 n+1}
	\nonumber\\&+\frac{4 \gamma ^4 (n-1)^{\frac{4}{n}-3} n^2 \left(n^3-2 n^2-n+3\right)}{4 n^2+8 n+3}+O\left(\gamma ^6\right) \Big), \label{thn2}\\
	w_{n-1}-w_{n}=&\pi  2^{3-2 n} (n-1)^{2 n+\frac{1}{n}-2} n^{2-2 n} \gamma ^{-2 n}\Big(
	\gamma +\frac{1}{2} \gamma ^3 (n-1)^{\frac{2}{n}-2} \left(3 n^2-6 n+2\right)\nonumber\\&
	+\frac{1}{8} \gamma ^5 (n-1)^{\frac{4}{n}-4} \left(15 n^4-80 n^3+144 n^2-104 n+24\right)+O\left(\gamma ^7\right)\Big).\label{thn3}
\end{align}
When $n=2$,  there is only one $\gamma$-parameter and $c_j$ can be solved exactly:
\begin{align}
	c_1=&\gamma\sum_{k=0}^\infty\frac{(-\gamma^2)^{k} \Gamma \left(k+\frac{1}{2}\right) (1+u)_k}{\Gamma (k+1) (\frac{3}{2}+u)_k}
	=\sqrt{\pi } \gamma  \, _2F_1\left(\frac{1}{2},u+1;u+\frac{3}{2};-\gamma ^2\right),\\
	c_2=&\sum_{k=0}^\infty\frac{(-\gamma^2)^{k} \Gamma \left(k+\frac{1}{2}\right) (-u)_k}{\sqrt{\pi } \Gamma (k+1) (-u+\frac{1}{2})_k}
	=\sqrt{\pi } \, _2F_1\left(\frac{1}{2},-u;\frac{1}{2}-u;-\gamma ^2\right),
\end{align}
where $(a)_{b}=\Gamma(a+b)/\Gamma(b)$ is the standard Pochhammer symbol.
One can check that all the undesired singularities of $\tilde f$ and  $\tilde X$  are removed. Using (\ref{coupling}) and (\ref{expwl1}) we get
\begin{align}
	\theta_{1}=&4 \tan ^{-1}\gamma,\\
	w_{1}=&\csc ^2\frac{\theta_1 }{2}+\frac{1}{4} (2\pi-\theta_1 )  \sin \theta_1  \csc ^4\frac{\theta_1  }{2},~~~
	w_{2}=\csc ^2\frac{\theta_1  }{2}-\frac{1}{4} \theta_1  \sin \theta_1   \csc ^4\frac{\theta_1  }{2},\label{n2w}
\end{align}
which are the same as the results obtain in \cite{Zarembo:2020tpf}.

Based on the result in the case of $n=2$, we assume that $w_{n-1}$ and $w_{n}$ take the form
\begin{equation}\label{nn1w}
	w_{n-1}=T^+_n(\theta_{n-1})+(2\pi-\theta_{n-1})T^-_n(\theta_{n-1}),
	~~~w_{n}=T^+_n(\theta_{n-1})-\theta_{n-1}T^-_n(\theta_{n-1}),
\end{equation}
where $T^+_n(\theta)=T^+_n(2\pi-\theta)$ and $T^-_n(\theta)=-T^-_n(2\pi-\theta)$.
From the expansions (\ref{thn1}), (\ref{thn2}) and (\ref{thn3}), one can try to conjecture the exact expressions of $w_{n-1}$ and $w_n$ as functions of $\theta_{n-1}$.
The difference $w_{n-1}-w_{n}$ can be conjectured to be
\begin{equation}
	w_{n-1}-w_{n}=2\pi T^{-}_n(\theta_{n-1})	=\pi  4^{1-n} (n-1) n \csc ^{2 n}\frac{\theta _{n-1}}{2}\sin \theta _{n-1}.
\end{equation}
For  $n=3,4,5,6$, $T^+_n(\theta)$ can be conjectured to be
\begin{align}
	T^+_3(\theta)=&\frac{1}{8} (\cos \theta +5) \csc ^4\left(\frac{\theta }{2}\right),\\
	T^+_4(\theta)=&-\frac{1}{160} (-14 \cos \theta +\cos 2 \theta -47) \csc ^6\frac{\theta }{2},\\
	T^+_5(\theta)=&\frac{(113 \cos \theta -13 \cos 2 \theta +\cos 3 \theta +319) \csc ^8\frac{\theta }{2}}{2688},\\
	T^+_6(\theta)=&-\frac{(-734 \cos \theta +106 \cos 2 \theta -14 \cos 3 \theta +\cos 4 \theta -1879) \csc ^{10}\frac{\theta }{2}}{43008}.
\end{align}
At this stage it is difficult to conjecture the expression of $T^+_n$ for general $n$.

\subsection{$n=3$}
When $n=3$,  we write $\tilde f$ and $\tilde X$ explicitly by using (\ref{eqal})-(\ref{eqan2}):
\begin{align}
	\tilde f=&\Big(a^+_1+a^+_3\frac{1}{2\cos\pi u -1}\gamma_2^{-u}\gamma_1^{-u}+a_{2}^+\frac{\sec\pi u}{2}\gamma_1^{-u},
	a^+_2+a^+_3\frac{1}{2\cos\pi u -1}\gamma_2^{-u},
	a^+_3\Big),\\
	\tilde X=&\Big(a^-_1,
	a^-_2-a_{1}^-\frac{\sec\pi u}{2} \gamma_1^{u},
	a^-_3-a^-_2\frac{1}{2\cos\pi u -1}\gamma_1^{u}-a_{1}^-\frac{\sec\pi u}{2} \gamma_1^{u}\gamma_2^{u}\Big),
\end{align}
where
\begin{align}
	a^{-}=&i\left(c_1 \frac{ \Gamma (u+1)}{\Gamma \left(u+\frac{3}{2}\right)},c_2\frac{  2^{1-2 u} 3^{3 u} \Gamma (u+1) \Gamma (2 u+1)}{\sqrt{\pi } \Gamma (3 u+2)},c_3 \frac{ \Gamma \left(\frac{3 u}{2}+\frac{1}{2}\right)}{\Gamma \left(\frac{3 u}{2}+1\right)}\right),\\
	a^{+}=&\left(c_1\frac{ \Gamma \left(-u-\frac{1}{2}\right)}{\Gamma (-u)},
c_2	\frac{ \sqrt{\pi } 2^{2-2 u} 3^{3 u} \Gamma (-3 u-1)}{\Gamma (-2 u) \Gamma (-u)},
	c_3\frac{ \Gamma \left(-\frac{1}{2} (3 u)\right)}{\Gamma \left(\frac{1}{2} (1-3 u)\right)}\right).
\end{align}

To remove undesired singularities of $\tilde f$ at $u=-(1+2m)/2$ and $u=-(1+2m)/3$ and $\tilde X$ at $u=(1+2m)/2$ and $u=(1+2m)/3$, $m=1,2,...$,
we make the ansatz 
\begin{align}
	c_1=&K^{-1}\frac{\left(u+\frac{1}{2}\right) \left(\sqrt[3]{2} \sqrt{\gamma_1} \sqrt[3]{\gamma_2}\right) }{\sqrt{3}}
	\sum _{m=0}^{\infty } \sum _{k=0}^{\infty } \sum _{j=0}^{\infty } \frac{\gamma_2^{\frac{2 k}{3}} \gamma_1^{j+m} f_1(m,k,j)}{m+u+\frac{1}{2}},\\
	c_2=&K^{-1}\frac{\left(3 \sqrt[3]{\gamma_2}\right) \left(u+\frac{1}{3}\right) }{2\ 2^{2/3}}
	\sum _{m=0}^{\infty } \sum _{k=0}^{\infty } \sum _{j=0}^{\infty }\left(\frac{ \gamma_2^{\frac{2 k}{3}} \gamma_1^{j+m+1} f_2(m,k,j)}{-m+u-\frac{1}{2}}
	+\frac{\gamma_1^j \gamma_2^{\frac{2 k}{3}+\frac{2 m}{3}} g_2(m,k,j)}{\frac{1}{3} (2 m+1)+u}\right),\\
	c_3=&K^{-1}+K^{-1}\sum _{m=0}^{\infty }\sum _{k=0}^{\infty } \sum _{j=0}^{\infty }  \frac{\gamma_1^j \gamma_2^{\frac{2 k}{3}+\frac{2 m}{3}+\frac{2}{3}} g_3(m,k,j)}{u-\frac{1}{3} (2 m+1)}.
\end{align}
Requiring the undesired singularities to be removable, we find the recurrence relations
\begin{align}
	g_{3}(m,k,j)=&\frac{(-9)^m 2^{\frac{1}{3} (-2) (5 m+2)} \sin \left(\frac{1}{3} \pi  (m-1)\right) \Gamma \left(\frac{2 (m+2)}{3}\right) \Gamma \left(\frac{1}{3} (4 m+5)\right)}{\pi  \Gamma (m+1)^2}\nonumber\\&\times
	\left(\sum _{p=0}^k \frac{3 g_2(p,k-p,j)}{2 (m+p+1)}-\sum _{p=0}^{j-1} \frac{6 f_2(p,k,j-p-1)}{-4 m+6 p+1}\right),\\
	g_{2}(m,k,j)=&\frac{(-9)^m 2^{-\frac{10 m}{3}} \cos \left(\frac{1}{6} (2 \pi  m+\pi )\right) \Gamma \left(\frac{2 m}{3}+\frac{1}{3}\right) \Gamma \left(\frac{4 m}{3}+\frac{2}{3}\right)}{\pi  \Gamma (m+1)^2}\nonumber\\&\times
	\left(\delta _{j,0} \delta _{k,0}+\sum _{p=0}^{k-1} -\frac{3 g_3(p,k-p-1,j)}{2 (m+p+1)}\right),\\
	f_{2}(m,k,j)=&-\frac{(-1)^m 2^{2 m+1} 3^{-3 m-2} \Gamma \left(3 m+\frac{5}{2}\right)}{\sqrt{\pi } \Gamma (m+1) \Gamma (2 m+2)}
	\sum _{p=0}^j \frac{f_1(p,k,j-p)}{m+p+1},\\
	f_{1}(m,k,j)=&-\frac{(-4)^m 3^{-3 m-1} \Gamma \left(3 m+\frac{3}{2}\right)}{\sqrt{\pi } \Gamma (m+1) \Gamma (2 m+1)}
	\nonumber\\&\times
	\left(\sum _{p=0}^{j-1} \frac{f_2(p,k,j-p-1)}{-m-p-1}+\sum _{p=0}^k \frac{6 g_2(p,k-p,j)}{-6 m+4 p-1}\right).
\end{align}
The overall factor $K$ is determined by $c_3(0)=3\sqrt{\pi } /2$. The relation between the $\theta$-parameters and $\gamma$-parameters can be obtained using (\ref{coupling}). For the first few orders in $\gamma_1$ and $\gamma_2$ we find
\begin{equation}
	\begin{split}
	\theta_1=&\sqrt[3]{\gamma _2} \left(\frac{4 \sqrt[3]{2} \sqrt{\gamma _1}}{\sqrt{3}}-\frac{8 \sqrt[3]{2} \gamma _1^{3/2}}{27 \sqrt{3}}+\frac{28 \sqrt[3]{2} \gamma _1^{5/2}}{243 \sqrt{3}}-\frac{416 \sqrt[3]{2} \gamma _1^{7/2}}{6561 \sqrt{3}}+O\left(\gamma _1^{9/2}\right)\right)
	\\&
	+\gamma _2 \left(-\frac{\sqrt{\gamma _1}}{\sqrt{3}}-\frac{16 \gamma _1^{3/2}}{27 \sqrt{3}}+\frac{32 \gamma _1^{5/2}}{243 \sqrt{3}}-\frac{400 \gamma _1^{7/2}}{6561 \sqrt{3}}+O\left(\gamma _1^{9/2}\right)\right)+O\left(\gamma _2^{5/3}\right),\\
		\theta_2=&\sqrt[3]{\gamma _2} \left(3 \sqrt[3]{2}-\frac{2 \sqrt[3]{2} \sqrt{\gamma _1}}{\sqrt{3}}+\frac{1}{3} \sqrt[3]{2} \gamma _1+\frac{4 \sqrt[3]{2} \gamma _1^{3/2}}{27 \sqrt{3}}-\frac{7}{81} \sqrt[3]{2} \gamma _1^2+O\left(\gamma _1^{5/2}\right)\right)\\&
		+\gamma _2 \left(-\frac{3}{4}+\frac{\sqrt{\gamma _1}}{2 \sqrt{3}}-\frac{7 \gamma _1}{12}+\frac{8 \gamma _1^{3/2}}{27 \sqrt{3}}+\frac{4 \gamma _1^2}{81}+O\left(\gamma _1^{5/2}\right)\right)
		+O\left(\gamma _2^{5/3}\right).
	\end{split}
\end{equation}
The normalized expectation values of the Wilson loop $w_l$ can be calculated using (\ref{expwl1}). The first few orders in  $\gamma_1$ and $\gamma_2$ are given by
\begin{equation}
	\begin{split}
		&w_3=\frac{1}{10}+\gamma _2^{2/3} \left(\frac{9}{35 \sqrt[3]{2}}+\frac{3}{35} 2^{2/3} \gamma _1-\frac{\gamma _1^2}{35 \sqrt[3]{2}}+\frac{2}{315} 2^{2/3} \gamma _1^3+O\left(\gamma _1^4\right)\right)\\&
		\phantom{w_3=}+\gamma _2^{4/3} \left(\frac{9}{35\ 2^{2/3}}+\frac{6}{35} \sqrt[3]{2} \gamma _1+\frac{1}{35} \sqrt[3]{2} \gamma _1^2-\frac{2}{315} \sqrt[3]{2} \gamma _1^3+O\left(\gamma _1^4\right)\right)+O\left(\gamma _2^{2}\right),\\
		&w_2-w_3=\frac{\frac{2 \sqrt[3]{2} \pi }{81}-\frac{2}{729} \left(\sqrt[3]{2} \pi \right) \gamma _1+\frac{8 \sqrt[3]{2} \pi  \gamma _1^2}{6561}-\frac{1144 \left(\sqrt[3]{2} \pi \right) \gamma _1^3}{1594323}+O\left(\gamma _1^4\right)}{\gamma _2^{5/3}}\\&
		\phantom{w_2-w_3=}+\frac{\frac{11 \pi }{162}+\frac{55 \pi  \gamma _1}{1458}-\frac{121 \pi  \gamma _1^2}{13122}+\frac{14201 \pi  \gamma _1^3}{3188646}+O\left(\gamma _1^4\right)}{\gamma _2}+O\left(\frac{1}{\gamma _2^{1/3}}\right),\\
		&w_1-\frac{w_3+w_2}{2}=
		\frac{\frac{\sqrt{3} \pi }{32\ 2^{2/3} \gamma _1^{3/2}}-\frac{\pi }{24 \left(2^{2/3} \sqrt{3}\right) \sqrt{\gamma _1}}-\frac{7 \pi  \sqrt{\gamma _1}}{432 \left(2^{2/3} \sqrt{3}\right)}+\frac{91 \pi  \gamma _1^{3/2}}{17496\ 2^{2/3} \sqrt{3}}+O\left(\gamma _1^{5/2}\right)}{\gamma _2^{5/3}}\\&
		\phantom{w_1-\frac{w_3+w_2}{2}=}+\frac{\frac{11 \sqrt{3} \pi }{256 \gamma _1^{3/2}}+\frac{11 \pi }{384 \sqrt{3} \sqrt{\gamma _1}}-\frac{517 \pi  \sqrt{\gamma _1}}{6912 \sqrt{3}}+\frac{11 \pi  \gamma _1^{3/2}}{2187 \sqrt{3}}+O\left(\gamma _1^{5/2}\right)}{\gamma _2}+O\left(\frac{1}{\gamma _2^{1/3}}\right).
	\end{split}
\end{equation}

\begin{figure}[H]
	\centering
	\includegraphics[width=0.6\linewidth]{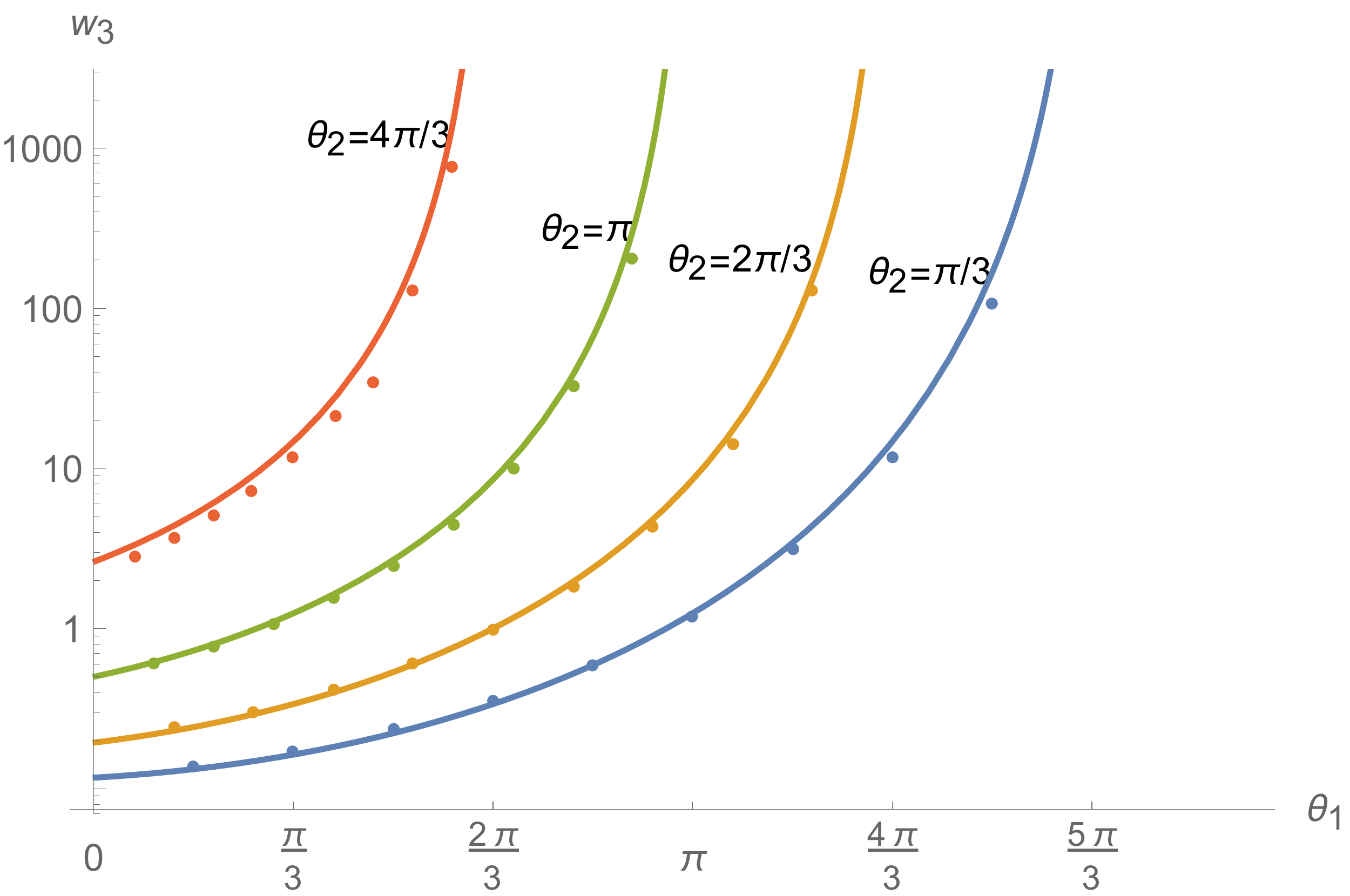}
	\caption{$w_3$ as a function of $\theta_1$ with different fixed $\theta_2$. The dots represent the data obtained by numerically solving the integral equations with $\lambda=1000$.}
	\label{fig:w3}
\end{figure}

From these expansions, the exact expression of $w_l$ as a function of the couplings can be conjectured to be
\begin{equation}\label{conjecture3}
	w_l=\frac{3}{16}\sum_{j=1}^3\csc^2\frac{\theta_j}{2}\csc^2\frac{\theta_{j+1}}{2}
	+\frac{3}{4}\sum_{j=1}^3\theta_{j}\csc^3\frac{\theta_l}{2}\csc^3\frac{\theta_{j}}{2}\sin\frac{\theta_{j}-\theta_{l}}{2}.
\end{equation}
We test the conjecture numerically and the result is plotted in fig. \ref{fig:w3}.
If this conjecture is true, we find
\begin{equation}
	w_l-w_j=
	\frac{3}{8}\csc^3\frac{\theta_l}{2}\csc^3\frac{\theta_{j}}{2}\sin\frac{\theta_{j}-\theta_{l}}{2}.
\end{equation}
Therefore the node with the largest 't Hooft coupling has the largest expectation value of Wilson loop.

\subsection{$n=4$ with two couplings}
\begin{figure}[H]
	\centering
	\includegraphics[width=0.7\linewidth]{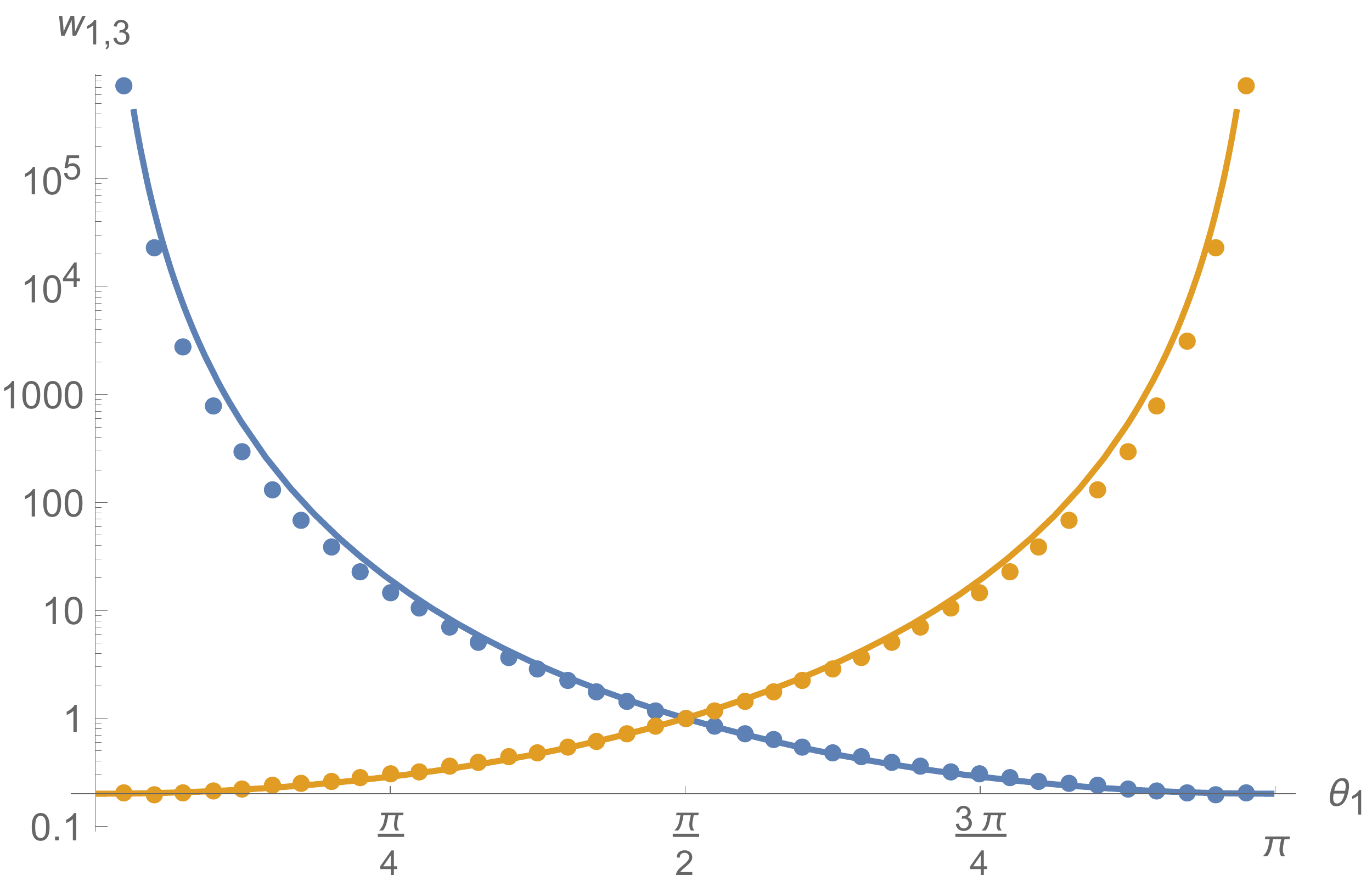}
	\caption{ A graph of $w_{1,3}$ as functions of $\theta_1$.  The dots represent the data obtained by numerically solving the integral equations with $\lambda=1000$.}
	\label{fig:w1133}
\end{figure}

It is a formidable task to deal with three independent $\gamma$-parameters in the 4-node case.
Here we consider a simple case where  $\lambda_1=\lambda_2$ and $\lambda_3=\lambda_4$.
We parameterize the 't Hooft couplings as
\begin{equation}
	\lambda_1=\lambda_2=\frac{\pi\lambda}{2\theta_1},~~~\lambda_3=\lambda_4=\frac{\pi\lambda}{2\pi-2\theta_1}.
\end{equation}
Because of the symmetry of the four-node quiver,  it is reasonable to expect $\eta_1=\eta_2$ and $\eta_3=\eta_4$ so  the factorization problem reduces to
\begin{equation}
	\begin{split}
		\left(
		\begin{array}{c}
			\tilde X_1   \\
			\tilde X_3  \\
		\end{array}
		\right) 
		=&
		\left(
		\begin{array}{cc}
			i \cot \pi  u-\frac{i}{2} \csc \pi  u & -\frac{i}{2} \csc \pi  u\gamma^{-3u} \\
			-\frac{i}{2} \csc \pi  u \gamma^{3u}& i \cot \pi  u-\frac{i}{2} \csc \pi  u \\
		\end{array}
		\right)
		\left(
		\begin{array}{c}
			\tilde f_1   \\
			\tilde f_3  \\
		\end{array}
		\right) \\
		=&\frac{2}{\sec \pi  u+2}M_3
		\left(
		\begin{array}{c}
			\tilde f_1   \\
			\tilde f_3  \\
		\end{array}
		\right) 
		.			
	\end{split}
\end{equation}
where $\gamma^3=\eta_1/\eta_3$ and $M_3$ is given by (\ref{Mn}). 
Therefore one can use the result in subsection \ref{ss32}.
Taking account into the extra scalar factor, we have
\begin{equation}
	\begin{split}
	f_1=&\frac{2^u 3^{-\frac{3 u}{2}-1} c_3(u) \gamma ^{-3 u} \Gamma \left(-\frac{u}{2}\right) \Gamma \left(\frac{1}{2}-u\right)}{\sqrt{\pi } (2 \cos (\pi  u)-1) \Gamma \left(\frac{1}{2}-\frac{3 u}{2}\right)}+\frac{\sqrt{\pi } 2^{u+3} 3^{\frac{3 u}{2}} c_2(u) \Gamma \left(1-\frac{u}{2}\right) \Gamma (-3 u-1)}{\Gamma \left(1-\frac{3 u}{2}\right) \Gamma (-u)^2},\\
	f_3=&\frac{3^{-\frac{3 u}{2}} 4^{-u} c_3(u) \Gamma \left(\frac{1}{2}-u\right) \Gamma \left(1-\frac{u}{2}\right) \Gamma \left(-\frac{1}{2} (3 u)\right)}{\pi  \Gamma (1-3 u)},	
	\end{split}
\end{equation}
where $c_{2,3}$ are given by (\ref{cncn}) with $n=3$. 
Expanding $f_l$ at small $u$ and solving the constraint \ref{bc}, we find
\begin{equation}
	\theta _1= \frac{4 c_2(0)}{3 \sqrt{\pi }}.
\end{equation}
The normalized Wilson loop expectation values are
\begin{equation}
	w_1=	\frac{9 c_3(-2) \gamma ^6}{10 \sqrt{\pi }}+\frac{8}{27} \sqrt{\pi } c_2(-2),~~~
	w_3=  \frac{9 c_3(-2)}{10 \sqrt{\pi }}.
\end{equation}
One can compute the first few terms in small $\gamma$ by using the recurrence relation (\ref{rec}).
The exact expression can be conjectured to be
\begin{equation}
	\begin{split}\label{conjecture1133}
		w_1=&\frac{1}{4} (\cos 2 \theta_1 +5) \csc ^4\theta_1 +\frac{3}{4} (\pi -\theta_1 ) \sin 2 \theta_1  \csc ^6\theta_1 ,\\
	w_3=&\frac{1}{4} (\cos 2 \theta_1 +5) \csc ^4\theta_1 -\frac{3}{4} \theta_1  \sin 2 \theta_1  \csc ^6\theta_1 .	
	\end{split}
\end{equation}
These results are plotted in fig. \ref{fig:w1133}.

\section{Nearly equal couplings}\label{s4}
In this section we consider the limit $\Delta_j \rightarrow 0$ so the couplings are expected to be nearly equal.
 We  also consider the limit $n \rightarrow \infty$ with $\Delta_j$ fixed. We find that the differences between the $\theta$-parameters are of order $n^{-2}$  and hence the couplings are also nearly equal.

\subsection{Small $\Delta$ limit} 
To solve the problem (\ref{sfx}) in the small $\Delta$ limit, it is convenient to decompose $\tilde f$ into a superposition of the eigenvectors of $\tilde{R}$  given by
\begin{equation}
	(v_j)_l=\exp (\frac{2\pi i}{n} jl-u x_l),
\end{equation}
where we denote $x_i=-\pi \Delta_i$. The eigenvalues are
\begin{align}
	r_j&=i \csc \pi  u \left(\cos \pi  u-\cos \frac{2 \pi  j}{n}\right)=\frac{r^-_j}{r^+_j},\\
	r_j^+&=\frac{2^{-u-1} \Gamma \left(-\frac{j}{n}-\frac{u}{2}+1\right) \Gamma \left(\frac{j}{n}-\frac{u}{2}\right)}{\Gamma (1-u)},\\
	r_j^-&=\frac{i \pi  2^{-u} \Gamma (u)}{\Gamma \left(-\frac{j}{n}+\frac{u}{2}+1\right) \Gamma \left(\frac{j}{n}+\frac{u}{2}\right)},
\end{align}
with $j=0,...,n-1$. We have  analytically factorized the eigenvalues.
When the  $x$-parameters are small and of the same order we propose an ansatz
\begin{equation}\label{fP}
	\tilde f= \sum_{j=0}^{n-1} P_j(u) r_j^+ v_j,
\end{equation}
where $ P_j(u) $ depends on the  variables $x_l$. The expansion of $P_j$ for small $x_l$ is
\begin{equation}
	P_j=P_j^{(0)}+P_j^{(1)}+P_j^{(2)}+O(x^3),
\end{equation}
where
\begin{equation}
	P_j^{(0)}(u)=\delta_{j0},
\end{equation}
and $P_j^{(1)}(u)$ and $P_j^{(2)}(u)$ are linear and quadratic in the $x$-parameters respectively.
This ansatz already assures the desired analytic properties of $\tilde{f}$ and $\tilde{X}$ except at $0$ and $\infty$.
Then the coefficients $P_j^{(n)}(u)$ are determined by the conditions (\ref{zero}) and (\ref{infinity}).
Expanding $r_j$ at small and large $u$, we find
\begin{align}
	r_j^+&\stackrel{u\rightarrow 0}{\simeq }
	\begin{cases}
		-\frac{1}{u}+\log (2)+O\left(u\right), &   j=0
		\\
		\frac{1}{2} \pi  \csc \left(\frac{\pi  j}{n}\right)+O\left(u\right), & \mathrm{otherwise}
	\end{cases},\\
	r_j^+&\stackrel{u\rightarrow -\infty}{\simeq  }
	\sqrt{\frac{\pi }{2}}(-u)^{-1/2}+\sqrt{\frac{\pi }{2}}\left(\frac{2 j^2}{n^2}-\frac{2 j}{n} +\frac{1}{4}\right)(-u)^{-3/2}+O((-u)^{-5/2}).
\end{align}
So $P_j^{(n)}$ should be a polynomial in $u$ of degree $n$ and
\begin{equation}
	P_j^{(n)}(0)=0,
\end{equation}
for $n\geq 1$. We find
\begin{align}
	P_j^{(1)}=&\frac{u}{n}  \sum _{m=1}^n e^{-\frac{2\pi i}{n}j m} x_m,\label{P1}\\
	P_j^{(2)}=&\sum _{m=1}^n e ^{-\frac{2\pi i}{n} j m} \left(-\frac{u \left(j^2-j n\right) x_m^2}{n^3}+\frac{u^2 x_m^2}{2 n}\right)\nonumber \\&
	+\sum _{p=0}^{n-1} \sum _{m_1=1}^{n} \sum _{m_2=1}^{n} \omega ^{\frac{2\pi i}{n}(-j m_2-p m_1 +p m_2)}\frac{2 u (j-p) (j-n+p)}{n^4} x_{m_1} x_{m_2}.
	\label{P2}
\end{align}
Expanding (\ref{fP}) at small $u$, we find
\begin{align}
	B_l=&\frac{x_l -P_0'(0)}{2 \pi },\label{Blsmall}\\
	C_l=&\frac{8 x_l P_0'(0)+8 S_l-4 x_l^2  -4 P_0''(0)-4 \log ^2(2)+\log ^2(4)}{32 \pi ^2},
\end{align}
with
\begin{equation}
	S_l=\sum_{j=1}^{n-1} \frac{1}{2} \pi  P_j'(0) e^{\frac{2 i \pi  j l}{n}} \csc \left(\frac{\pi  j}{n}\right).
\end{equation}
The constraint (\ref{bc}) becomes
\begin{equation}
	\sum_{j=0}^{n-1}4 i \pi  e^{\frac{2 i \pi  j \left(l+\frac{1}{2}\right)}{n}} \sin ^2\left(\frac{\pi  j}{n}\right) P_j'(0)
	=\frac{2\pi}{n}  \theta _{l+1}-\frac{2\pi}{n}   \theta _l,
\end{equation}
or equivalently
\begin{equation}\label{thx}
	\theta _l=\frac{2\pi}{n}+\sum_{j=0}^{n-1} \frac{4}{n}   e^{\frac{2 i \pi  j l}{n}} \sin \left(\frac{\pi  j}{n}\right)P_j'(0).
\end{equation}

To compute the strong-coupling expectation values of Wilson loops, we evaluate $\tilde f$ at $u=-2$:
\begin{equation}
	\tilde f_l(-2)=\sum _{j=0}^{n-1} \frac{(n-j)j}{n^2} \pi   P_j(-2)  \csc \left(\frac{\pi  j}{n}\right) e^{2 x_l +\frac{2 i \pi  j l}{n}}.
\end{equation}
Then equation (\ref{ratiow}) leads to
\begin{equation}\label{ratio}
	w_l=\sum _{j=0}^{n-1}\pi  \frac{(n-j)j}{n^2}  \csc \left(\frac{\pi  j}{n}\right) e^{2 P_0'(0)+\frac{2 i \pi  j l}{n}} P_j(-2). \\
\end{equation}
When the $x_l$'s are small, up to the first nontrivial order we have
\begin{equation}
	P_j(u)=\delta_{j0}+	\frac{u}{n}  \sum _{m=1}^n e^{-\frac{2\pi i}{n}j m} x_m+O(x^2).
\end{equation}
We denote $\alpha_l=\theta _l-2\pi/n$ as the fluctuation of $\theta_l$ around $2\pi/n$. Using (\ref{thx}) and (\ref{ratio}) we find
\begin{equation}\label{smallw}
	w_l=
	1-\frac{\pi }{2n^2}\sum _{j=1}^{n-1} \sum _{l=1}^n j (n-j)  \csc ^2\left(\frac{\pi  j}{n}\right) \cos \left(\frac{2 \pi  j (m-l)}{n}\right)\alpha_l+O(\alpha^2).
\end{equation}
When all the $\alpha$-parameters go to zero, we get $w_l\rightarrow 1$ as expected.

To check the conjecture (\ref{conjecture3}), we compute $w_l$ up to the second nontrivial order in the $n=3$ case. Without loss of generality  we set $x_1+x_2+x_3=0$ because only differences between the $x$-parameters appear in the kernel $R$.
Using (\ref{P1}) and (\ref{P2}) we get
\begin{equation}
	\begin{split}
		P_0(u)=&1+\frac{1}{27}  u (9 u+8) \left(x_1^2+x_2 x_1+x_2^2\right)+O\left(x^3\right),\\
		P_1(u)=&-\frac{1}{6} i  u \left(\left(\sqrt{3}-3 i\right) x_1-\left(\sqrt{3}+3 i\right) x_2\right)\\&
		+\frac{1}{108} u (9 u+4) \left(\left(1-i \sqrt{3}\right) x_1^2+4 x_2 x_1+\left(1+i \sqrt{3}\right) x_2^2\right)+O\left(x^3\right),\\
		P_2(u)=&\frac{1}{6} i   u \left(\left(\sqrt{3}+3 i\right) x_1-\left(\sqrt{3}-3 i\right) x_2\right)\\&
		+\frac{1}{108} u (9 u+4) \left(\left(1+i \sqrt{3}\right) x_1^2+4 x_2 x_1+\left(1-i \sqrt{3}\right) x_2^2\right)+O\left(x^3\right).\\
	\end{split}
\end{equation}
Substituting into (\ref{P1}) and solve $x_1$ and $x_2$ in small $\alpha_l$ leads to
\begin{equation}
	\begin{split}
		x_1=&\frac{\sqrt{3}\alpha _1}{2}-\frac{1}{18} \left(\alpha _1^2-2 \alpha _2 \alpha _1-2 \alpha _2^2\right)+O\left(\alpha^3\right) ,\\
		x_2=&\frac{\sqrt{3}\alpha _2}{2}+\frac{1}{18} \left(2 \alpha _1^2+2 \alpha _2 \alpha _1-\alpha _2^2\right)+O\left(\alpha^3\right) .
	\end{split}
\end{equation}
Using (\ref{ratio}) we find
\begin{equation}
	\begin{split}
		w_1=&1-\frac{4}{9} \left(\pi  \alpha _1\right) +\left(\frac{2 \pi  \left(\alpha _1^2-2 \alpha _2 \alpha _1-2 \alpha _2^2\right)}{9 \sqrt{3}}+ \left(\alpha _1^2+\alpha _2 \alpha _1+\alpha _2^2\right)\right)  +O\left(\alpha^3\right),\\
		w_2=&1-\frac{4}{9} \left(\pi  \alpha _2\right) +\left(\frac{2 \pi  \left(-2 \alpha _1^2-2 \alpha _2 \alpha _1+\alpha _2^2\right)}{9 \sqrt{3}}+ \left(\alpha _1^2+\alpha _2 \alpha _1+\alpha _2^2\right)\right) +O\left(\alpha^3\right) ,\\
		w_3=&1+\frac{4}{9} \pi  \left(\alpha _1+\alpha _2\right) +\left(\frac{2 \pi  \left(\alpha _1^2+4 \alpha _2 \alpha _1+\alpha _2^2\right)}{9 \sqrt{3}}+ \left(\alpha _1^2+\alpha _2 \alpha _1+\alpha _2^2\right)\right) +O\left(\alpha^3\right) ,
	\end{split}
\end{equation}
which are consistent with the conjecture (\ref{conjecture3}) expanded around $\theta_l=2\pi/n$.

\subsection{Large $n$ limit}
\begin{figure}[H]
	\centering
	\includegraphics[width=0.6\linewidth]{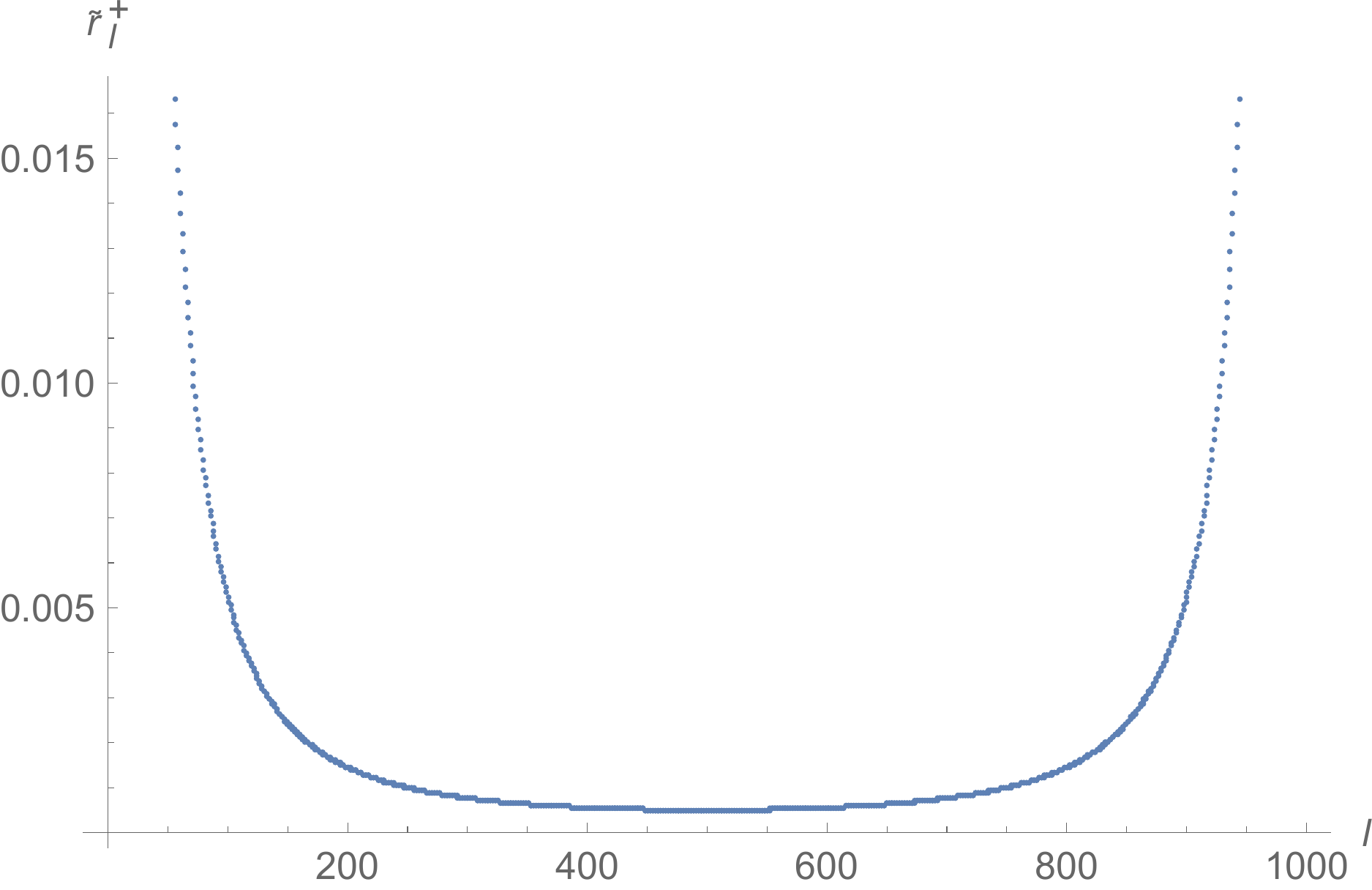}
	\caption{$\tilde r_{l}^+(-2)$ with $n=1000$.}
	\label{fig:rl1000}
\end{figure}

The limit of large $n$ is interesting because it is related to the dimensional deconstruction \cite{ArkaniHamed:2001ca,Hill:2000mu}.
In this case, the theory becomes effectively five-dimensional and the quiver can be interpreted as the fifth dimension.
We require that the $x $-parameters stay finite and  vary slowly along the quiver, i.e.  $ x_{j+1}- x_j\sim O(n^{-1})$ .  
To compute higher order corrections to $P_j$, we make the ansatz
\begin{equation}
	P_j(u)=\frac{1}{n}\sum _{m=1}^n e^{-\frac{2\pi i}{n}j m} e^{u x_m } Q(u,\frac{m}{n}),
\end{equation}
where $Q$ is a function to be determined. Then we have
\begin{equation}
	\begin{split}
		\tilde f_l&= \frac{1}{n}\sum _{m=1}^n Q(u,\frac{m}{n}) \tilde r_{l-m}^+
		e^{u x_m -u x_l }.\\
		\tilde r_l^+&=\sum_{j=0}^{n-1} e^{\frac{2\pi i}{n}j l} r_j^+ .
	\end{split}
\end{equation}
Using the asymptotics of $ r_j^+$ at infinity
\begin{equation}
	\begin{split}
		\sqrt{\frac{2}{\pi }}r_j^+=&
		(-u)^{-1/2}+(-u)^{-3/2} \left(\frac{2 j (j-n)}{n^2}+\frac{1}{4}\right)+\frac{(-u)^{-5/2} \left(8 j^2-8 j n+n^2\right)^2}{32 n^4}\\&
		+O\left((-u)^{-7/2}\right),
	\end{split}
\end{equation}
we find
\begin{equation}
	\begin{split}
		\frac{1}{n}\sqrt{\frac{2}{\pi }}\tilde r_l^+=&
		(-u)^{-1/2}\delta _{l0}+(-u)^{-3/2} \frac{ \left(12 \csc ^2\left(\frac{\pi  l}{n}\right)-\delta _{l0} \left(12 \csc ^2\left(\frac{\pi  l}{n}\right)+n^2-4\right)\right)}{12 n^2}\\&
		+(-u)^{-5/2}\frac{\left(60 \left(1-\delta _{l0}\right) (n^2-(n^2+8) \cos (\frac{2 \pi  l}{n})-16) \csc ^4\left(\frac{\pi  l}{n}\right)+\left(7 n^4+40 n^2-32\right) \delta _{l0}\right)}{480 n^4}\\&
			+O\left((-u)^{-7/2}\right).
	\end{split}
\end{equation}
In the limit of large $n$ with $s\equiv l/n$ fixed, we get
\begin{equation}
	\begin{split}
		\frac{1}{n}\sqrt{\frac{2}{\pi }}\tilde r_l^+=&
		(-u)^{-1/2}\delta _{l0}+(-u)^{-3/2}\left(-\frac{\delta_{l0}}{12}+\frac{\csc ^2(\pi  s)+\left(\frac{1}{3}-\csc
			^2(\pi  s)\right) \delta_{l0}}{n^2}+O\left(n^{-4}\right)\right)\\&
		+(-u)^{-5/2}\left(\frac{7 \delta_{l0}}{480}-\frac{\csc ^2(\pi  s) \left((\cos (2 \pi s)+5) \delta _{l0}-6\right)}{24 n^2}+O\left(n^{-4}\right)\right)
		\\&
				+O\left((-u)^{-7/2}\right).
	\end{split}
\end{equation}
For finite $s$,  the $n^{-2}$ corrections are neglectable, but when $s \rightarrow 0$ or $s \rightarrow 1$ these corrections can be large. It is reasonable to assume $\tilde r_{l}^+\simeq 0$ to all order of the large $u$ expansion when $s$ is not close to 0 or $1$. To test this assumption, we compute $\tilde r_{l}^+(-2)$ with $n=1000$ and the result is shown in fig. \ref{fig:rl1000}.

Therefore if we view $s$ as a continuous variable with the identification $s\sim s+1$, for $u$ on the negative real axis $\tilde r_l^+(u)$ acts on slowly varying $x_l$ like a Dirac delta function on a circle:
\begin{equation}\label{rs}
	\tilde r_{l}^+(u)\simeq r_0^+(u) \delta(s),~~~s=\frac{l}{n},
\end{equation}
 where 
\begin{equation}
	\frac{1}{n}\sum_{l=1}^n \tilde r_{l}^+(u)=r_0^+(u).
\end{equation} 
has been taken into account. It is worth mentioning that (\ref{rs}) is not valid for general $u$. For instance, when $u\rightarrow 0$ we have 
\begin{equation}
	\tilde r_{l}^+(u) \simeq \sum_{j=0}^{n-1} e^{\frac{2\pi i}{n}j l} 
	(\frac{\delta_{j0}}{u}+O(u^0)) =\frac{1}{u}+O(u^0),
\end{equation}
In the case of $u<0$, we have
\begin{equation}
	\begin{split}
		\tilde f_{l}&\simeq \int_0^1 ds_1 Q(u,s)\delta(s-s_1) r_0^+(u)
		e^{u x(s_1) -u x(s) }\\
		&= Q(u,s)r_0^+(u),
	\end{split}
\end{equation}
where $x(l/n)\equiv x_{l}$.
When $Q$ is independent of $u$ we get the desired asymptotics at infinity. Taking into account the small $u$ expansion, we find $Q(u,s)=1$.
So $\tilde f$ at negative real $u$ does not depend on the $x$-parameters in the large $n$ limit, so we have
\begin{equation}
	P_j(u)\simeq\frac{1}{n}\sum _{m=1}^n e^{-\frac{2\pi i}{n}j m} e^{u x_m }.
\end{equation}
Equation (\ref{thx}) becomes
\begin{equation}
	\alpha_l=\theta _l-2\pi/n
	\simeq \frac{4}{n^2} \sum_{j=0}^{n-1}\sum _{m=1}^n    e^{\frac{2 i \pi  j (l-m)}{n}} \sin \left(\frac{\pi  j}{n}\right)
	x_m.
\end{equation}
The  right hand side are of order $n^{-2}$, so  $\alpha_l\sim O(n^{-2})$.
Therefore the 't Hooft couplings
\begin{equation}
	\lambda_l=\frac{2\pi\lambda}{n \theta}=\frac{2\pi\lambda}{2 \pi+n \alpha_l}\simeq\lambda(1+O(n^{-1})),
\end{equation}
are nearly equal in this limit.

Solving the $x$-parameters in terms of the $\alpha$-parameters, we get
\begin{equation}
	\begin{split}
		x_m -\frac{1}{n}\sum_{l=1}^n x_l&\simeq 
		\frac{1}{4}\sum_{j=1}^{n-1}\sum _{l=1}^n \csc \left(\frac{\pi  j}{n}\right) e^{\frac{2 i \pi  j (m-l)}{n}}\alpha_l	\\
		&\simeq \frac{1}{2}\sum_{j=1}^{n/2}\sum _{l=1}^n \frac{n}{\pi j} \cos  \left(\frac{2 \pi  j (m-l)}{n}\right)\alpha_l.
	\end{split}
\end{equation}
Replacing the discrete variables by continuous functions, we find
\begin{equation}
	x(s)-\int_{0}^1 ds' x(s')
	=	-\int_{0}^1 ds' \frac{\log \left(4 \sin ^2(\pi ( s-s'))\right)}{2 \pi }\hat\alpha(s'),
\end{equation}
where $\hat\alpha $ is defined as
\begin{equation}
\alpha_{l}=	n^{-2}\hat\alpha(l/n),
\end{equation}
and we have used
\begin{equation}
	\sum _{j=1}^{\infty } \frac{\cos (2 \pi  j s)}{\pi  j}=-\frac{\log \left(4 \sin ^2(\pi  s)\right)}{2 \pi }.
\end{equation}
Using (\ref{Blsmall}) and $\tilde f_l(-2)\simeq 1$, the positions of the endpoints and the expectation values of Wilson loops are given by
\begin{align}
	\mu_l
	\simeq&\frac{\sqrt{\lambda}}{2\pi}+
	\int_{0}^1 ds' \frac{\log \left(4 \sin ^2(\pi ( s-s'))\right)}{2 }\hat\alpha(s'),\\
	w_l \simeq &\exp\left(	\int_{0}^1 ds' \frac{\log \left(4 \sin ^2(\pi ( s-s'))\right)}{\pi }\hat\alpha(s') \right) .\label{largenw}
\end{align}

\section{A conjecture}\label{s5}
Based on the results above, we can actually make the following conjecture:
\begin{equation}\label{wn}
	w_i=T(\theta)+\sum_{j=1}^{n} (w_i-w_j)\frac{\theta_j}{2\pi},
\end{equation}
where
\begin{align}
	w_i-w_{i+1}=&\pi  2^{2-2 n} n\left(\sum_{j=0}^{n-2}\sin \left(\frac{\sum _{k=0}^j (\theta_{i+k+1}-\theta_{i-k})}{2}\right) \csc \left(\frac{\sum _{k=0}^j \theta_{i+k+1}}{2}\right) \csc \left(\frac{\sum _{k=0}^j \theta_{i-k}}{2}\right)\right)\label{wd}\nonumber\\&\times
	\prod _{j=0}^{n-2} \csc \left(\frac{\sum _{k=0}^j \theta_{i-k}}{2}\right) \csc \left(\frac{\sum _{k=0}^j \theta_{i+k+1}}{2}\right),\\
	T(\theta)=&2^{2 - 2 n} n\sum_{i=1}^{n}\prod _{j=0}^{n-2} \csc \left(\frac{\sum _{k=0}^j \theta_{i-k}}{2}\right) \csc \left(\frac{\sum _{k=0}^j \theta_{i+k+1}}{2}\right).
\end{align}
Note that (\ref{wn}) is consistent because
\begin{equation}
w_i-w_j= (w_i-w_j)\sum_{k=1}^{n}\frac{\theta_k}{2\pi}	=\sum_{k=1}^{n} (w_i-w_k)\frac{\theta_k}{2\pi}-\sum_{k=1}^{n} (w_j-w_k)\frac{\theta_k}{2\pi}
.
\end{equation}
One can first compute the differences between any $w_i$ and $w_j$ using (\ref{wd}), and then $w_i$ using (\ref{wn}).
This conjecture is consistent with the hierarchical coupling limit results (\ref{largenw}) and (\ref{nn1w}), $n=2$ result (\ref{n2w}), $n=3,4$ conjectures (\ref{conjecture3}) and (\ref{conjecture1133}), and the nearly equal coupling limit result (\ref{smallw}).

The conjecture (\ref{wn}) implies that the  Wilson loops depend on the  $\theta$-parameters almost analytically.
As discussed in \cite{Rey:2010ry} non-analytic behavior emerges when one of the $\theta$-parameters approaches zero.
 The Wilson loop $w_l$ diverges as $\theta_l^{-3}$ when $\theta_l\rightarrow 0$.
 
One can try to understand the conjecture  in the context of the brane construction\cite{Hanany:1996ie,Witten:1997sc}.
The starting point is Type IIA theory with a compactified $x^6$ direction.
There are $n$ NS5-branes extended in the 12345 directions on the $x^6$ circle
and also $N$ D4-branes wrapped around $x^6$ circle and stretched between the  NS5-branes. 
The $n$-node quiver gauge theory describes the low-energy dynamics of these D4-branes.
The parameter $\theta_i$ is proportional to the distance between the $(i-1)$th and $i$th NS5-branes in the $x^6$ direction.
Therefore the term $ \csc \left(\sum _{k=0}^j \theta_{i+k+1}/2\right)$ appearing in the conjecture
 has a natural  geometric interpretation as the inverse of the chordal distance between  the $i$th and $(i+j)$th NS5-branes.

\begin{figure}[H]
	\centering
	\includegraphics[width=0.48\linewidth]{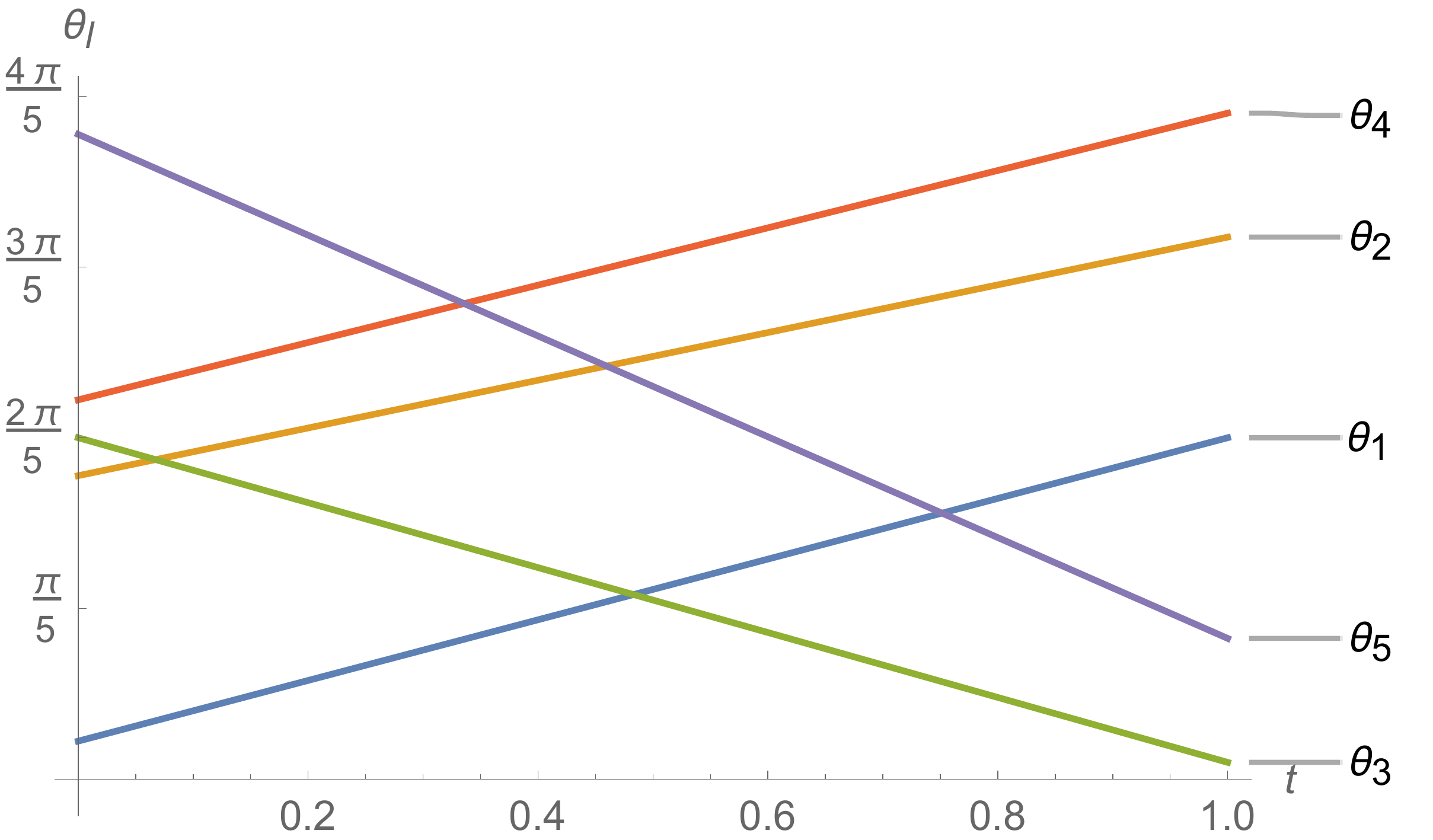}
	\includegraphics[width=0.48\linewidth]{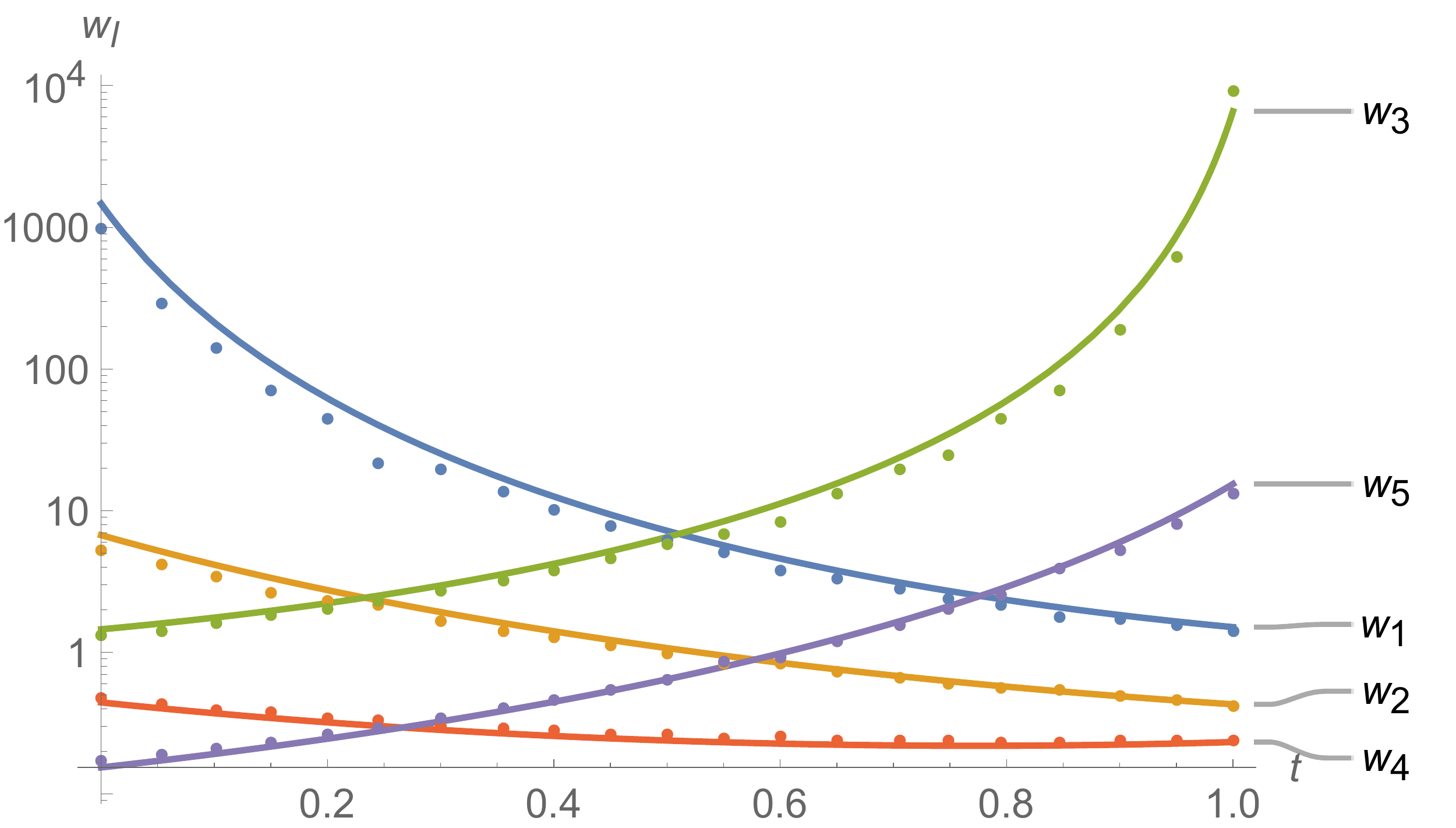}
	\caption{ The $\theta$-parameters and the normalized Wilson loop expectation values on the curve (\ref{curve1}).  The dots represent the data obtained by numerically solving the integral equations with $\lambda=1000$.}
	\label{fig:n=5}
\end{figure}

To test the conjecture numerically, we consider the $n=5$ case. There are four independent $\theta$-parameters.
We choose a curve in the parameter space:
\begin{equation}\label{curve1}
	\theta_j=\frac{2}{5} \pi  \left(1+\frac{1}{9} (i-3)^3 (1-t)+t \sin \left(\frac{4}{5} \pi  (i-1)\right)\right).
\end{equation}
where $t$ is a parameter restricted to the interval [0,1].
On this curve expectation values of Wilson loops agree well with numerics (fig. \ref{fig:n=5}).

\begin{figure}[H]
	\centering
	\includegraphics[width=0.7\linewidth]{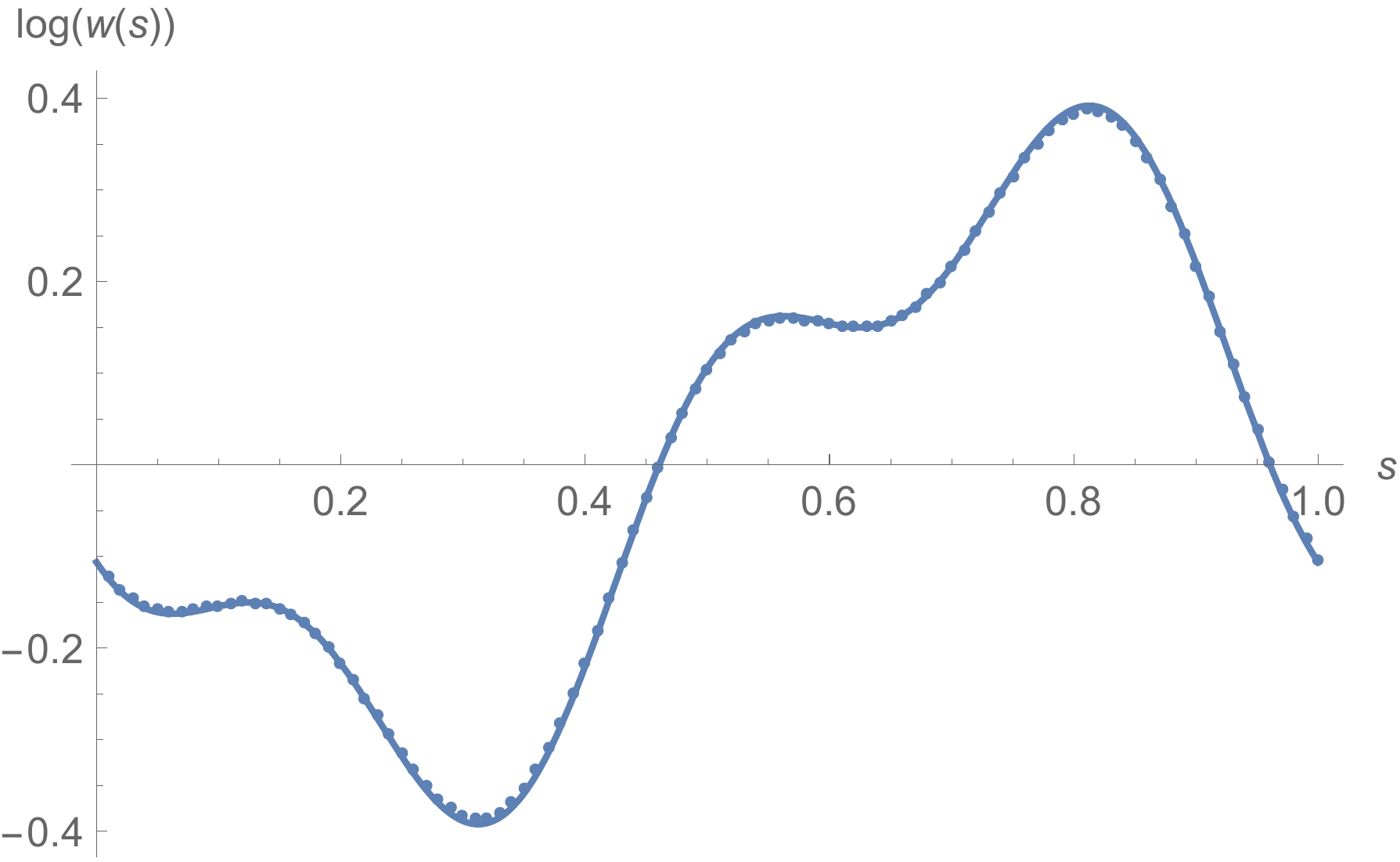}
	\caption{ The curve depicts  $\log w(s)=-\frac{\sin (2 \pi  s)}{\pi }-\frac{\cos (6 \pi  s)}{3\pi  }$ obtained from the large $n$ approximation (\ref{largenw}) with the $\alpha$-parameters given by (\ref{alpha1}). The dots represent the same quantity obtained from (\ref{wn}) with $n=100$.}
	\label{fig:largenw}
\end{figure}

Finally we compare the large $n$ approximation (\ref{largenw}) with the conjecture (\ref{wn}).
To check that  $\alpha_l\sim O(n^{-2})$ leads to a well-defined limit, we choose the values of $\alpha_l$ to be 
\begin{equation}
	\begin{split}\label{alpha1}
\alpha_l=	\frac{2 \sin \left(\frac{2 \pi  l}{n}\right)}{n^2}+\frac{2 \cos \left(\frac{6 \pi  l}{n}\right)}{n^2}.
	\end{split}
\end{equation}
Plugging in the conjecture (\ref{wn}) with different $n$, we find
\begin{equation}
	\begin{split}
	n=10:&~~ \log w_{1}\approx -0.139445,\\
n=50:&~~ \log w_{1}\approx -0.131851,\\
n=100:&~~ \log w_{1}\approx -0.120574.		
	\end{split}
\end{equation}
Therefore one can see that $\log w_{1}\sim O(n^0)$ if  $\alpha_l\sim O(n^{-2})$ .
With the $\alpha$-parameters given by (\ref{alpha1}),  the large $n$ formula (\ref{largenw}) gives
\begin{equation}
	\log w(s)=-\frac{\sin (2 \pi  s)}{\pi }-\frac{\cos (6 \pi  s)}{3\pi  }.
\end{equation}
This agrees very well with the result obtained from (\ref{wn}) (fig. \ref{fig:largenw}).

It would also be interesting to consider the limit of large $n$  with  finite $\lambda_l/\lambda\sim O(n^0)$.  By applying the conjecture (\ref{wn}) to some explicit examples, we find $\log w_{1}\sim O(n)$  and equation (\ref{largenw}) is no longer valid.

\section{Conclusions}\label{sc}

In this paper, we investigated the expectation values of the circular Wilson loops in the $n$-node
superconformal quiver theories at strong coupling using localization.
We computed the Wilson loops  expectation values when the couplings are hierarchically different or nearly equal, and 
finally, made a conjecture for arbitrary strong couplings. It would be interesting to derive the conjecture (\ref{wn}) rigorously.
One possible way is to relate the  matrix factorization problem to a solvable quantum system as the $n=2$ case in \cite{Zarembo:2020tpf}. 

To compare our result with the string prediction, one needs to compute the one-loop correction to the string worldsheet integral in a similar way as was done in $AdS_5\times S^5$ \cite{Forini:2015bgo,Faraggi:2016ekd,Forini:2017whz,Cagnazzo:2017sny,Medina-Rincon:2018wjs}.
In the orbifold case, the $\theta$-dependence is expected to arise from the worldsheet instanton  wrapping the collapsed two-cycles.
We hope the comparison can be achieved  in the nearby future.

\section*{Acknowledgments}
I would like to thank Rob Klabbers and Konstantin Zarembo  for collaboration
at the initial stage of this work and enlightening discussions.
I  also thank  Konstantin Zarembo  for providing the code for numerical verification.
I thank Jun-Bao Wu and Konstantin Zarembo  for valuable comments on the draft. 
I thank Marco Bill\`{o}, Nadav Drukker, Francesco Galvagno, and Alberto Lerda for useful discussions.
This work was supported by the grant ``Exact Results in Gauge and String Theories'' from the Knut and Alice Wallenberg foundation.

%\appendix

\bibliographystyle{jhep}  
\bibliography{quiver}

\end{document}